\pdfoutput=1 
\documentclass[runningheads]{llncs}

\usepackage{orcidlink}
\usepackage{color}
\usepackage{rotating}
\usepackage{color}
\usepackage{graphicx}
\usepackage{svg}
\usepackage{textcomp}
\usepackage{xurl}
\usepackage{tabularx}
\usepackage{pgf}
\usepackage{layouts}
\usepackage{geometry}
\usepackage{optidef}
\usepackage{soul}
\usepackage{color}
\usepackage{tabularx}
\usepackage{graphicx}%
\usepackage{multirow}%
\usepackage{mathrsfs}%
\usepackage{pdflscape}
\usepackage{svg}
\usepackage{subcaption}

\usepackage[abs]{overpic}
\usepackage{pict2e}
\usepackage{tikz}
\usetikzlibrary{shapes.geometric}
\definecolor{ao}{rgb}{0.0,0.5,0.0}
\definecolor{aoblue}{rgb}{0.0549, 0.3608, 0.6549}
\definecolor{afb}{rgb}{0.36,0.54,0.66}
\definecolor{pinkpink}{rgb}{1,0,1}
\definecolor{blueblue}{rgb}{0,0,1}
\definecolor{bluebluelight}{rgb}{0.0392,0.3529,0.6549}
\definecolor{otherblue}{rgb}{0.1215,0.4666,0.7058}


\begin{document}

\title{The Dark Side of Flexibility: How Aggregated Cyberattacks Threaten the Power Grid}

\author{Daniel Myrén\inst{1} \and
Zeeshan Afzal\inst{2} \and
Mikael Asplund\inst{2}}

\institute{Sectra Communications, Linköping, Sweden \and
Linköping University, Linköping, Sweden }

\maketitle

\begin{abstract}

Flexible energy resources are increasingly becoming common in smart grids. These resources are typically managed and controlled by aggregators that coordinate many resources to provide flexibility services. However, these aggregators and flexible energy resources are vulnerable, which could allow attackers to remotely control flexible energy resources to launch large-scale attacks on the grid. This paper investigates and evaluates the potential attack strategies that can be used to manipulate flexible energy resources to challenge the effectiveness of traditional grid stability measures and disrupt the first-swing stability of the power grid. Our work shows that although a large amount of power is required, the current flexibility capacities could potentially be sufficient to disrupt the grid on a national level.

\end{abstract}

\section{Introduction}
\label{cha:introduction}

\begin{figure*}[!htb]
   \centering
    \includegraphics[width=0.80\textwidth]{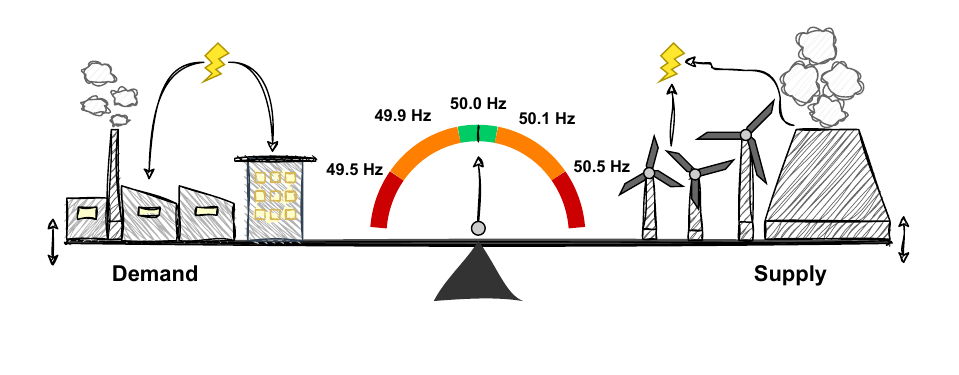}
    \caption{Balance between supply and demand and its impact on frequency}
     \label{figure:balance}
\end{figure*}

The electric grid must always maintain a balance between demand and supply~\cite{security-bess} to ensure continuous and efficient power delivery. This balance is typically measured through the frequency of alternating current (AC) and is set at 50 Hz in the EU (also called nominal frequency). Figure~\ref{figure:balance} illustrates this task of maintaining balance. When power production and consumption are perfectly balanced, the grid frequency stays at its nominal value. If consumption exceeds production, the frequency decreases. If production exceeds consumption, the frequency increases. Consequently, the grid operators continuously monitor the frequency to detect and mitigate any imbalances. If the frequency deviations are not promptly managed and counteracted, there is a risk of significant disruptions such as blackouts and damage to critical infrastructures such as generators~\cite{iva_svangmassa}. 

The growing prevalence of renewable energy sources (RESs) such as solar and wind in the power grid makes power generation more variable and unpredictable, creating a challenge to maintain the balance between supply and demand~\cite{problem_renewable_load_demand}. This increased variability creates a need for flexible energy resources (FERs) such as energy storage systems and heat pumps which are increasingly utilized by grid operators as a solution to maintain the stability of the grid. These resources can quickly adapt their energy consumption or inject more energy into the grid to restore balance when required, and are expected to grow significantly in numbers over the coming years \cite{flex_report}. Another significant challenge with a higher share of RESs and FERs is that most RESs do not inherently contribute to grid inertia. As a result, a grid with reduced inertia and higher variability in energy production becomes more sensitive and susceptible to fluctuations in demand and bidirectional energy flow, where certain FERs, such as batteries, can both consume power from the grid (charging) and supply energy back to it (discharging). Unfortunately, these factors introduce new risks and vulnerabilities to the power grid, increasing its exposure to both cyber and physical threats. Adversaries could manipulate FERs to increase their demand or inject energy into the grid, to execute large-scale distributed attacks with potentially catastrophic consequences~\cite{dabrowski_load_altering}. Furthermore, the ability to remotely control these resources enables attackers to launch large scale attacks capable of impacting the stability of the power grid~\cite{security-bess}. By targeting aggregators, attackers can potentially take control over and manage hundreds or thousands of distributed assets simultaneously.

As the capacity, prevalence, and reliance on FERs continue to grow, they will likely draw increased attention from adversaries. It is therefore crucial to investigate the threats posed by adversaries who can control and manipulate multiple FERs within the power grid. In this work, we identify potential attack methods or strategies where adversaries can manipulate demand and bidirectional energy flow of aggregated FERs and assess the impact of these strategies on the grid frequency stability. Moreover, we evaluate the severity of these attacks in the context of both current and future capacities of FERs and analyze the effectiveness of existing frequency mitigation measures such as the Fast Frequency Reserve (FFR) and Frequency Containment Reserves (FCR) in countering such threats. By investigating potential attack strategies, their impact, and their severity, we aim to provide insights into how FERs can be manipulated to affect the frequency stability of the power grid. This knowledge is expected to inform the development of robust defensive strategies to protect the power grid against these emerging threats. 

The remainder of this paper is structured as follows. Section~\ref{cha:background} provides necessary background
and an overview of the related work. It also highlights the gaps that we aim to fill. Section~\ref{cha:agg_att_sce} defines aggregated attack strategies that adversaries could employ to destabilize the grid. Section~\ref{cha:evaluation} describes the simulation environment, grid model, and results from evaluating the impact of identified attacks on the stability of the grid. Section~\ref{cha:discussion} provides a discussion on the results and outlines future research directions. Finally, section~\ref{cha:conclusion} provides conclusions.

\section{Background \& Related Work}
\label{cha:background}

The frequency of the power grid directly influences the speed of generators and other rotating components connected to the power grid \cite{iva_svangmassa}. If the frequency falls below 49.5 Hz, it becomes necessary to start disconnecting specific loads to reduce demand. Should the frequency, despite these measures, fall below 47.5 Hz, large power stations are automatically disconnected, potentially cascading to total system collapse and widespread power blackouts. Conversely, an increase in frequency is also problematic and dangerous for the grid. If the frequency exceeds 52 Hz, the power stations are disconnected from the grid to prevent damaging vital components. Regardless of whether the frequency increases or decreases from the nominal frequency,  exceeding these critical thresholds leads to the disconnection of loads and generators, potentially triggering a chain reaction that can ultimately cause the entire system to collapse. System-wide power blackouts would significantly disrupt and put enormous stress on society as critical infrastructure and essential services cease normal operations. This has been seen is several recent real-world incidents such as the cascading power outage in Puerto Rico, the major blackout in Spain and Portugal leaving millions without power for a half-day, and the attack against Ukraine's power grid~\cite{un_ukraine}. 

To maintain grid balance and correct any deviations in frequency, constant coordination is required among many actors, including power grid operators (TSOs and DSOs), energy producers, and consumers~\cite{iva_svangmassa}. In the Nordic countries, the power grids are tightly interconnected with bilateral connections~\cite{nordic_grid_plan}. The frequency is managed by coordinating the production and consumption within the Nordic synchronous grid, with each country incorporating frequency reserve resources that provide up and down regulation to minimize frequency fluctuations. Fluctuations in the power grid frequency are traditionally mitigated by two primary mechanisms: inertia provided by the large rotating masses such as turbines and frequency reserve resources~\cite{iva_svangmassa}. The rotational masses in turbines and generators directly connected to the grid act as an initial buffer against changes in demand and generation. If the frequency decreases, the synchronous generators rotate slower, releasing some of their rotational energy to the power grid. Conversely, if the frequency increases, the generators rotate faster, absorbing some of the excess energy in the grid as rotational energy. As RESs increasingly constitute a larger share of power generation, the grid's inertia is decreasing since RESs are typically not connected via synchronous generators. Consequently, this buffer is slowly vanishing, making the grid more susceptible to sudden changes in demand and supply, and requiring larger capacities of frequency reserves and FERs to manage fluctuations in frequency. 

\subsection{The Swedish Power Grid}

In Sweden, various frequency reserve resources are utilized to mitigate sudden frequency deviations and maintain the balance between demand and supply~\cite{iva_svangmassa,nordic_grid_plan}. These resources must have sufficient capabilities, including capacity and responsiveness, to stabilize the power grid during sudden changes in demand and supply. There are currently six types of frequency reserve resources, as listed in Table \ref{table:fcr}, along with their activation range, response time, and capacity \cite{svk_reserves}. As soon as a frequency deviation is detected, automatic frequency reserves are activated to perform up- or down-regulation to keep the frequency within the normal operating range of $49.90-50.10$ Hz \cite{iva_svangmassa}. According to a 2023 report~\cite{flex_report}, the estimated flexibility capacity of various FERs in Sweden is expected to reach 1747 MW in 2025 and increasing substantially to 8000 MW by 2030. Key contributors include large-scale battery systems (1200 MW by 2030), heat pumps (1300 MW), EV charging (500 MW), and hydrogen production via electrolysis, which is expected to grow from 120 MW in 2025 to a dominant 4400 MW in 2030. The increase reflects a strategic push to meet Sweden’s forecasted power demand of 25800 MW by 2030 (up from 17800 MW in 2023).

\begin{table*}[h!]

\caption{Frequency reserve and restoration products.}
\label{table:fcr}
\begin{center}    
     \small
    \begin{tabular}{|p{4cm}|p{2cm}|p{2cm}|p{2cm}|p{2cm}|}
    \hline
    \textbf{Reserve Type} & \textbf{Regulation Direction} & \textbf{Activation Range (Hz)} & \textbf{Response Time} & \textbf{Capacity (MW)} \\ \hline 
    Fast Frequency Reserve (FFR) & Upwards & Below 49.7-49.5 & 0.7-1.3 seconds & Up to 100\\ \hline 
    Upward Frequency Containment Reserve (FCR) - Disturbance & Upwards & Below 49.9   & Within seconds  & Up to 567\\ \hline
    Downward FCR - Disturbance & Downwards & Above 50.1 & Within seconds & Up to 547 \\ \hline 
    FCR - Normal & Both & Outside 50.0 & Within seconds & 235\\ \hline
    Automatic Frequency Restoration Reserve (aFRR) & Both & Outside 50.0 & Within 5 Minutes & Up to 111\\ \hline
    Manual Frequency Restoration Reserve (mFRR) & Both & Manual Activation & Within 15 Minutes & Up to 300\\ \hline
    
    \end{tabular}
\end{center}
\end{table*}

\label{previousincidents}

Unfortunately, there have been several instances in Sweden and worldwide where the grid frequency has deviated significantly from the nominal value. In Sweden, during 2011, the sudden disconnection of the Oskarshamn 3 nuclear plant, which supplied 1400 MW, caused the grid frequency to drop to 49.36 Hz within approximately ten seconds. According to the Swedish TSO, this is one of the lowest frequencies recorded within the N-1 criterion, which requires the grid to withstand the loss of the largest single generator (up to 1450 MW)~\cite{svk_integrering_2013}. The grid is designed to handle disturbances up to this magnitude, and the frequency reserves are dimensioned to support this criterion. During this critical period, frequency reserves were automatically activated, and the frequency recovered to 49.8 Hz within 25 seconds, successfully preventing a more severe grid failure. Another incident in February 2023, involved the Nordlink HVDC connection unexpectedly importing 300 MW instead of exporting 1375 MW, creating a 1675 MW imbalance within the power grid \cite{nordic_grid_plan}. The frequency briefly dropped to 49.41 Hz before recovering to the normal operating range. This event underscored the increasing risk posed by modern interconnections, as the potential imbalance in such failures can exceed traditional design limits like the N-1 criterion. A more severe event occurred in April 2023, when maintenance failures caused voltage drops and automatic disconnection of multiple generators, including Forsmark 1 and 2, removing 2130 MW from the grid. The frequency quickly dropped to 49.3 Hz within eight seconds. Emergency injections through HVDC connections and activation of frequency reserves (600 MW and 700 MW, respectively) helped restore stability to 49.8 Hz. Over the next minutes, the frequency restoration reserve (mFRR) added 2431 MW, and full normalization was achieved. Backup measures, including 7600 MW of additional mFRR and automatic load disconnection at 48.8 Hz, were available but not needed. The Swedish TSO concluded that despite the severity, the system remained within operational limits and the response measures were effective.

\subsection{Related Work}

A significant amount of prior work exists on smart grid security, covering a wide range of vulnerabilities and mitigation strategies. Research has demonstrated that the attack surface on the grid is large, and includes utility operators, aggregators, communication networks, smart meters, inverters, and energy management systems that ultimately manage and control various energy resources \cite{1_der_smart_inverters}. 

One significant challenge is insecure communication between autonomous cyber-physical systems, such as batteries and EVs, and their control systems, which often relies on legacy protocols designed without robust security measures~\cite{nist_smart_grid}. These protocols frequently lack encryption and authentication, making them vulnerable to attacks like eavesdropping and false message injection~\cite{scada_issues}. Such attacks can distort state estimations and disrupt grid operations~\cite{commmunication_grid}, highlighting the critical need to ensure the integrity of measurement data and control commands for safe and reliable system performance. Another major vulnerability lies within the Advanced Metering Infrastructure (AMI) which depends on real-time measurements from producers and remote control commands over the AMI network~\cite{security_issue_ami}. These remote commands could potentially be exploited allowing attackers to disconnect millions of connected smart meters \cite{security_issue_ami}. Similar concerns are raised by Parks \cite{parks_ami} and Li et al. \cite{smartgrid_attacks_challenges} who note that AMI components located in customer premises are susceptible to tampering, potentially allowing users to manipulate pricing  data for personal gain. 

Beyond AMI, vulnerabilities extend to surrounding infrastructure and control systems. Firewalls can be undermined by misconfigurations that permit unauthorized access \cite{security_grid_state_of_art}. Energy management systems and aggregators also present risks, as they often rely on insecure or outdated communication protocols and commercial off-the-shelf (COTS) components with internet-facing interfaces \cite{nist_smart_grid,scada_issues,smartgrid_attacks_challenges}. These systems, managing large numbers of distributed energy resources (DERs), become high-value targets for cyberattacks \cite{1_der_smart_inverters}. For example, Baumgart et al. \cite{inspection_bess} revealed vulnerabilities that allow remote access, eavesdropping, and unauthorized system changes \cite{inspection_bess}. Collectively, these vulnerabilities demonstrate that many systems within the smart grid could be exposed to large-scale coordinated attacks capable of disrupting numerous energy assets simultaneously.

Attacks on the power grid are increasingly sophisticated and target various components to disrupt the critical balance between supply and demand. False data injection (FDI) attacks could corrupt state estimation process by manipulating meter data, leading to wrong control decision and potential grid instability \cite{liu_false_data_injection,lin_false_data_injection,soltan_load_alt_sim}. Load-altering attacks, such as those evaluated by Dabrowski et al.~\cite{dabrowski_load_altering}, statically or dynamically~\cite{amini_load_alt_sim} change electricity demand of remotely controlled assets, such as air conditioning and electrical heating. \cite{mohsenian_load_altering}. These attacks can cause frequency deviations, trip tie lines, and even split grid zones. EVs and charging stations have also emerged as potential vectors for such attacks~\cite{acharya_load_alt_ev,sayed_v2g_attacks}. In addition, attackers can exploit bidirectional energy flow enabled by technologies like battery energy storage systems (BESSs) and Vehicle-to-Grid (V2G) to inject power into the grid at unscheduled times, causing frequency oscillations and generator disconnections~\cite{sayed_v2g_attacks,denai_v2g_impact}. Aggregated attacks further escalate the threat by coordinating malicious actions across multiple grid components, such as generators and transmission lines, to induce cascading failures and maximize disruption~\cite{xiang_aggregated,1_der_smart_inverters}. As grid digitalization expands, the integration of IT systems, IoT devices, and decentralized control mechanisms makes it increasingly feasible for attackers to coordinate large-scale cyber-physical attacks with potentially catastrophic consequences.

Despite extensive research, gaps remain in understanding how different aggregated attack strategies impact the grid frequency, particularly in the context of existing frequency mitigation measures. While static and dynamic load-altering attacks have been widely discussed, their definitions and implementations vary. This paper formalizes these attack types and introduces a structured classification that includes novel attacks that combine multiple attack strategies. Furthermore, existing research seldom evaluates attacks based on realistic flexibility capacities. This work evaluates attacks using different timings and aggregated attack sizes that reflect the current and future flexibility potential, identifying which FERs pose the highest risk based on their capacity. 

\section{Aggregated Attacks}
\label{cha:agg_att_sce}

Aggregated attacks involve load-altering attacks where attackers manipulate the demand or bidirectional energy flow of multiple FERs. By aggregating many FERs, attackers can execute large-scale attacks involving significant amounts of power. If a sufficient number of FERs are aggregated, the attack could violate frequency stability parameters discussed earlier. We build upon previously discussed concepts of static and dynamic load-altering attacks and formalize them into a structured classification. This classification introduces four aggregated attack categories: static, switching, periodic, and combination attacks, that could be used by adversaries to destabilize the grid. While the static, switching, and periodic attack concepts are loosely based on existing literature, the combination attacks are a novel contribution of this work.

\subsection{Static Attacks}
Static attacks aim to create a persistent imbalance between supply and demand through various attack strategies. These strategies can be further divided into four types: 

\begin{itemize}
    \item \textbf{Demand Increase (DI) Attack}: This attack targets unidirectional loads primarily drawing energy from the grid. By increasing the demand of controlled assets the frequency in the grid declines. 
    
    \item \textbf{Demand Reduction (DR) Attack}: Similar to the DI attack, this type typically involves unidirectional loads. However, contrary to a DI attack, this type decreases the energy demand of controlled assets, resulting in an increased grid frequency. 
    
    \item \textbf{Supply Increase (SI) Attack}: This attack typically involves assets capable of bidirectional energy flow, such as V2G and batteries. By simultaneously injecting additional power into the grid, the controlled assets cause the grid frequency to increase. This attack type is sometimes also referred to as a power injection attack.    

    \item \textbf{Supply Reduction (SR) Attack}: Like the SI attack, this type primarily involves bidirectional energy resources. However, controlled assets either reduce their power output or completely disconnect from the grid, resulting in a drop in grid frequency.
    
\end{itemize}

\begin{figure}[h]
  \centering
    \includegraphics[width=0.60\linewidth]{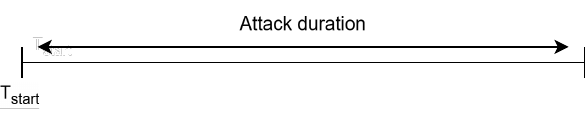}
    \caption{Timeline of a static attack}
    \label{figure:staticattack}
\end{figure}

These attacks require an attacker to have control over assets with sufficient capacity. The assets may either be concentrated in a specific grid location or distributed across a larger geographic area. Ultimately, once the attacker has achieved control over assets, they can execute a static attack. To do this, the attacker simultaneously executes one of the attack types described above at a predetermined start time, $T_{start}$, and sustains the attack indefinitely. This attack is relatively simple to execute since it only involves a single activation action. Figure~\ref{figure:staticattack} shows an illustration.

\subsection{Switching Attacks}
\begin{figure}[h]
  \centering
    \includegraphics[width=0.60\linewidth]{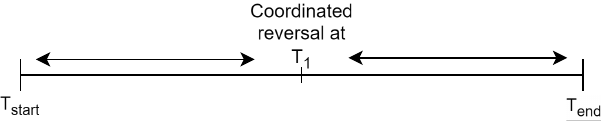}
    \caption{Timeline of a switching attack}
    \label{figure:switchingattack}
\end{figure}

A switching attack, depicted in Figure \ref{figure:switchingattack}, is an extension of a static attack. Unlike static attacks, which maintain an increased or reduced demand or supply, switching attacks include a sudden reversion of the attack, restoring the balance between supply and demand to their original state. This restoration does not occur for static DI, DR, SI, and SR attacks. The switching attack begins with executing a static attack at $T_{start}$, followed by a reversion at $T_1$. Once the attack is reverted and the balance is restored, it is considered complete. 

\subsection{Periodic Attacks}
\label{sec:periodicattacks}
Periodic attacks involve repeatedly executing a series of switching attacks against the grid at certain intervals with the aim of achieving sustained destabilization. We define periodic attacks to only involve one static attack type (DI, DR, SI, or SR) to allow for a more targeted evaluation. The timing of the switching attacks could be either predetermined or dynamically selected based on real-time grid conditions. To dynamically determine this timing, an attack could, for example, continuously monitor the grid's frequency to launch subsequent attacks at moments that cause maximum disruptions. These subsequent attacks could have the largest impact if executed when the slope of the frequency deviation is the steepest. 

\begin{figure}[h]
  \centering
    \includegraphics[width=0.60\linewidth]{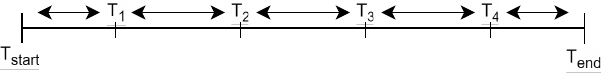}
    \caption{Timeline of a periodic attack}
    \label{figure:periodicattack}
\end{figure}

Figure \ref{figure:periodicattack} illustrates a periodic attack consisting of five switching attacks. The first switching attack occurs between  $T_{start}$ and $T_{1}$, followed by subsequent switching attacks between the determined intervals. The periodic attack can continue as long as the assets remain connected to the grid, can adjust their energy demand, or have sufficient charge to execute the attacks effectively.

\subsection{Combination Attacks}
A combination attack builds upon the previously mentioned attacks and incorporates a combination of demand-based (increasing or decreasing) and supply-based (increasing or decreasing) attacks. By alternating between attack strategies that increase the frequency (e.g. DR and SI attacks) and those that decrease the frequency (e.g. DI and SR attacks), an attacker could take advantage of and leverage the frequency response in the grid to maximize the impact of the attack. 

\begin{figure}[h]
  \centering
    \includegraphics[width=0.60\linewidth]{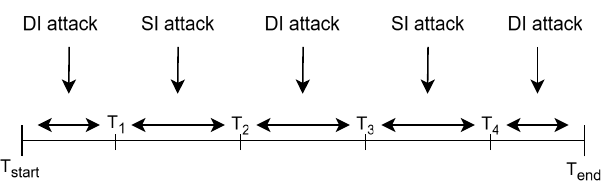}
    \caption{Timeline of a combination DI and SI attack}
    \label{figure:smartattack}
\end{figure}

Particularly, if an attacker manages to activate the grid's frequency reserve, it can perform subsequent attacks that leverage the upward or downward regulation. For instance, an attacker could first execute a static DI attack to reduce the grid frequency, triggering the frequency reserve to compensate by increasing supply. Following this, the attacker could perform an SI attack, further increasing the supply and creating an even larger imbalance between supply and demand. This example is illustrated in Figure \ref{figure:smartattack}, where the attacker executes the DI attack at $T_{start}$, followed by a SI at a strategically selected moment $T_1$, and then alternating between these.

\section{Evaluation}
\label{cha:evaluation}

This section presents an evaluation of the impact of the aggregated attack strategies. We begin by describing the employed simulation environment and the grid model. The impact of the attacks on the first-swing stability of the power grid model is then assessed by measuring the grid's frequency response to the attacks. These results are then put in the context using the predicted flexibility potential and frequency reserves capacity.

\subsection{Simulator and Model}

Several simulators are used in the literature to simulate a grid. These include Matlab Simscape \cite{simscape}, Siemens Power System Simulation \cite{siemenspsse}, Ge Vernova Steady State Power Flow \cite{gevernova}, and PowerWorld Simulator \cite{powerworld}. While most offer free versions for research, these are often limited to models with fewer than 50 buses and may lack full features. In this work, we use the PowerWorld simulator, which meets the necessary technical and accessibility requirements. It supports up to 40 buses and includes essential tools for analyzing first-swing stability and executing predefined scenarios.

Apart from the simulator, the selection of the power grid model is also vital because it determines the grid's behavior, including its reaction to changes in demand and supply. There is a large body of works that are based on two grid models, the \emph{WSCC 9-bus} \cite{wscc9bus}, a literature-based model, and the \emph{Hawaii 37-bus} \cite{hawaii37bus}, a synthetic model. The WSCC 9-bus model, consisting of three generators and nine buses, is a simplified theoretical model often used in research. Each bus is treated as a large substation, representing a broad geographical area. In contrast, the Hawaii 37-bus model, based on publicly available data from Oahu, includes 45 generators, 27 loads, and incorporates both conventional and renewable energy sources modeled with GENROU and REGC\_A machine models. To assess model accuracy and select a model for our work, both models were tested using simulated attack scenarios involving an 8\% increase in demand, designed to replicate the frequency response observed during the Oskarshamn 3 incident described in Section~\ref{cha:background}. Both models dropped below the 49.5 Hz frequency stability threshold, but the WSCC model stabilized more effectively and replicated the frequency response of the Swedish grid. The Hawaii model showed poorer stability and did not align with the observed data. Thus, due to its realistic and consistent frequency response, the WSCC 9-bus model is selected for this work. It offers a reliable platform for evaluating attack scenarios with reduced variability and dependence on specific load configurations.

Obviously, the WSCC 9-bus model cannot exactly represent the Swedish national grid in terms of complexity or magnitude. However, we use it as a proxy model in which both attacks and responses are scaled according to characteristics of the Swedish grid. The model behavior is also validated against real observed events to ensure that it is sufficiently representative. This allows us to gauge high-level system behavior and the potential impact of large-scale coordinated attacks. 

\subsection{Results}
An evaluation of aggregated attack strategies using simulations conducted in the PowerWorld simulator with the WSCC 9-bus model is presented below.

\subsection*{Static Attacks}
Static attacks are performed by executing a DI, DR, SI or SR attack on the grid and sustaining over a certain period of time. Although the controlled and manipulated assets can be distributed across many loads and injection points, we assume that they are concentrated on a single load. The following assesses the impact of static attacks on the grid frequency. 

\subsubsection*{Demand Increase Attacks}
Three DI attacks, where demand is increased by different amounts, are executed to evaluate the impact and behavior. The increments in demand are derived from past incidents described in Section~\ref{cha:background}. The incident at Oskarshamn 3 caused an imbalance of 1400 MW, the Nordlink incident caused an imbalance of 1675 MW, and the Forsmark 1 \& 2 incident caused an imbalance of 2130 MW. When compared to the average annual generation for the respective years, these incidents correspond to 8\%, 9.4\%, and 12\% of total generation in the grid, respectively. By applying the same demand changes observed in these previous incidents, we compare and benchmark the grid's frequency response based on the WSCC model against the actual incidents. 

\begin{figure}[!htb]
    \centering
    \begin{overpic}[scale=1.0,unit=1mm, width=0.69\textwidth,clip, trim=0cm 0.3cm 0.3cm 0cm]{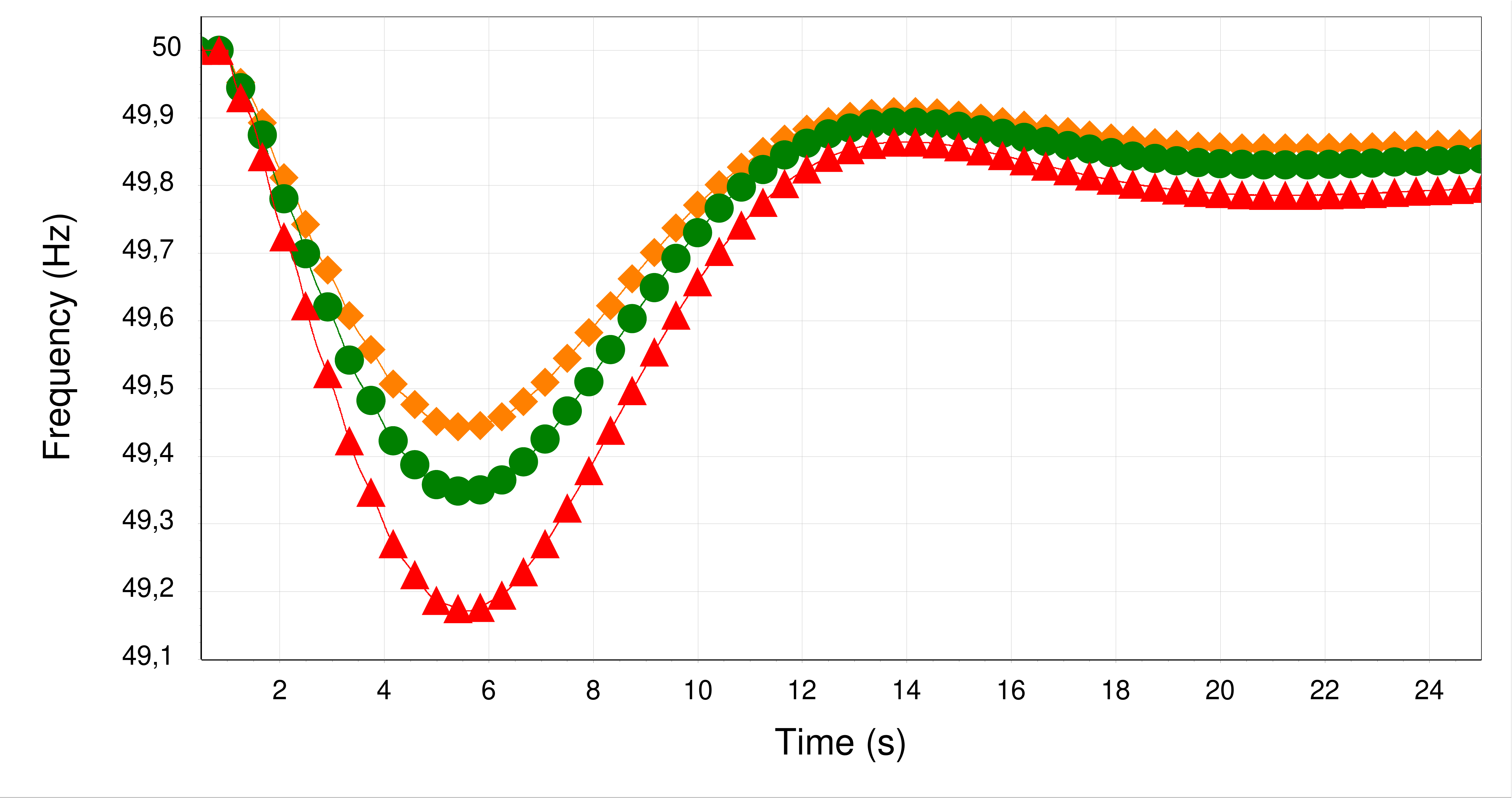}
        \put(60,30){\tikz\draw[rotate around={45:(0,0)},orange,fill=orange] (0,0) rectangle (0.2,0.2); 8\% increased demand}
        \put(60,25){\tikz\draw[ao,fill=ao] (0,0) circle (.7ex); 9.4\% increased demand}
        \put(60,20){\tikz\draw[red,fill=red] (0,0) -- (0.22,0) -- (0.11,0.22) -- cycle; 12\% increased demand}
    \end{overpic}
    \caption{Frequency response under three DI attack scenarios derived from real-world incidents.}
    \label{figure:lca_static_bus8}    
\end{figure}
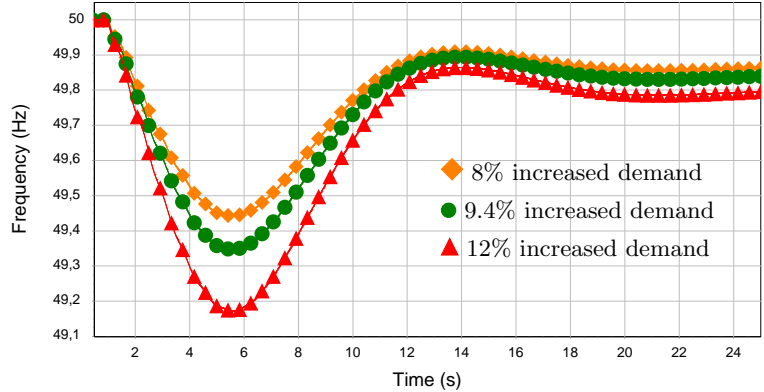

The results of the DI attacks with demand increases of 8\%, 9.4\%, and 12\% are illustrated in Figure \ref{figure:lca_static_bus8}. This clearly demonstrates that DI attacks that induce greater increases in demand lead to larger frequency deviations from the nominal frequency. A greater increase leads to a larger imbalance between supply and demand, resulting in a more significant frequency response.

Comparing these frequency responses observed in Figure \ref{figure:lca_static_bus8} against the actual incidents is important to assess how the frequency response in the model varies due to different demand increases. For the WSCC model, a static DI attack by 12\% (Forsmark 1 \& 2 incident) causes the grid's frequency to drop to 49.17 Hz before stabilizing at 49.8 Hz. In the actual incident, the lowest measured frequency was 49.3 Hz, stabilizing at 49.8 Hz. These results indicate that the WSCC model appears slightly more sensitive to DI attacks. However, this can be partly explained by the fact that, in the real incident, the frequency was 50.1 Hz before rapidly declining \cite{svk_storning_2023}. This difference results in a smaller frequency deviation from the nominal frequency (50 Hz), but when measured from the highest to lowest frequency, the frequency responses are similar. The frequency of the simulated grid in response to the other two DI attacks closely matches the frequency responses observed in the real incidents, with similar first-swing stability. However, the time in which this plays out does not accurately reflect the reality. For example, the frequency responses in Figure \ref{figure:lca_static_bus8} stabilize after approximately 12 seconds, while in the actual incidents, the time it takes for the Swedish grid to stabilize is around 30 seconds. This suggests that the simulator underestimates the time aspect, and the attacks may take longer for an actual grid to recover from. However, the lowest frequency and first-swing stability of the WSCC model are very similar to the Swedish grid, which is the primary focus of this paper. These DI attacks are sufficient to activate upwards-regulating frequency reserves, including FCR-Normal, FFR, and upwards FCR, as the frequency falls below 49.5 Hz. In addition, certain loads can be automatically disconnected to return the frequency to acceptable operating ranges. Consequently, an attacker that has the capability to execute attacks of this magnitude would pose a significant threat to the stability of the grid. In this case, the grid returns to within the normal operating range (49.9-50.1 Hz) but the sudden frequency drop to under 49.5 Hz could still result in deterioration of frequency reserves, disconnection of loads, or even further cascading instabilities. 

Furthermore, we note that the observed imbalances of 1400, 1675, and 2130 MW respectively were sufficient to cause a frequency drop below 49.5 Hz. Therefore, DI attacks replicating these incidents appear to be feasible in 2025 as they require control over a capacity of less than 1747 MW, which is the estimated flexibility potential of FERs in Sweden in 2025, as discussed in Section~\ref{cha:background}. Although such an attack would require considerable coordination and control of many FERs, the capacity and deployment of FERs is continuously growing, increasing the likelihood of an attacker controlling sufficient resources to perform such attacks in the coming years.

We conducted additional simulations with varying attack capacities to further analyze the impact of DI attacks on the grid. These are aimed to determine how different amounts of power impact the grid frequency. The results demonstrate a negative linear trend, indicating that the lowest frequency decreases as the attack capacity increases. Based on this data, a linear function ($Y=-0.06X+49.92$) is derived where $Y$, represents the lowest achieved frequency in Hz, as a function of $X$, the attack capacity expressed as a share of the total average generation in Sweden. This function can be used to approximate the lowest frequency induced by a DI attack.

\subsubsection{Demand Reduction and Supply Increase Attacks}

In terms of frequency response, DR and SI attacks have the opposite effect of DI attacks. While DI attacks increase the power demand of FERs, DR attacks reduce the energy demand, and SI attacks inject additional power into the grid. Both DR and SI attacks create a similar imbalance between supply and demand, causing the frequency in the grid to increase. During the attack simulation, it is not possible to increase the generator's power output. Therefore, SI attacks are modeled by applying a negative demand to a load, which results in a similar frequency response. However, DR and SI attacks may differ in other aspects that could be important when analyzing, such as voltage or phase stability, two other important stability measurements for the grid. For simplicity, the paper will only refer to DR attacks in the following. However, it should be noted that SI attacks are also included in the analysis and are attacks that can cause the frequency to increase, similar to DR attacks.

\begin{figure}[!htb]
    \centering
    \begin{overpic}[scale=1.0,unit=1mm, width=0.69\textwidth,clip, trim=0cm 0.3cm 0.3cm 0cm]{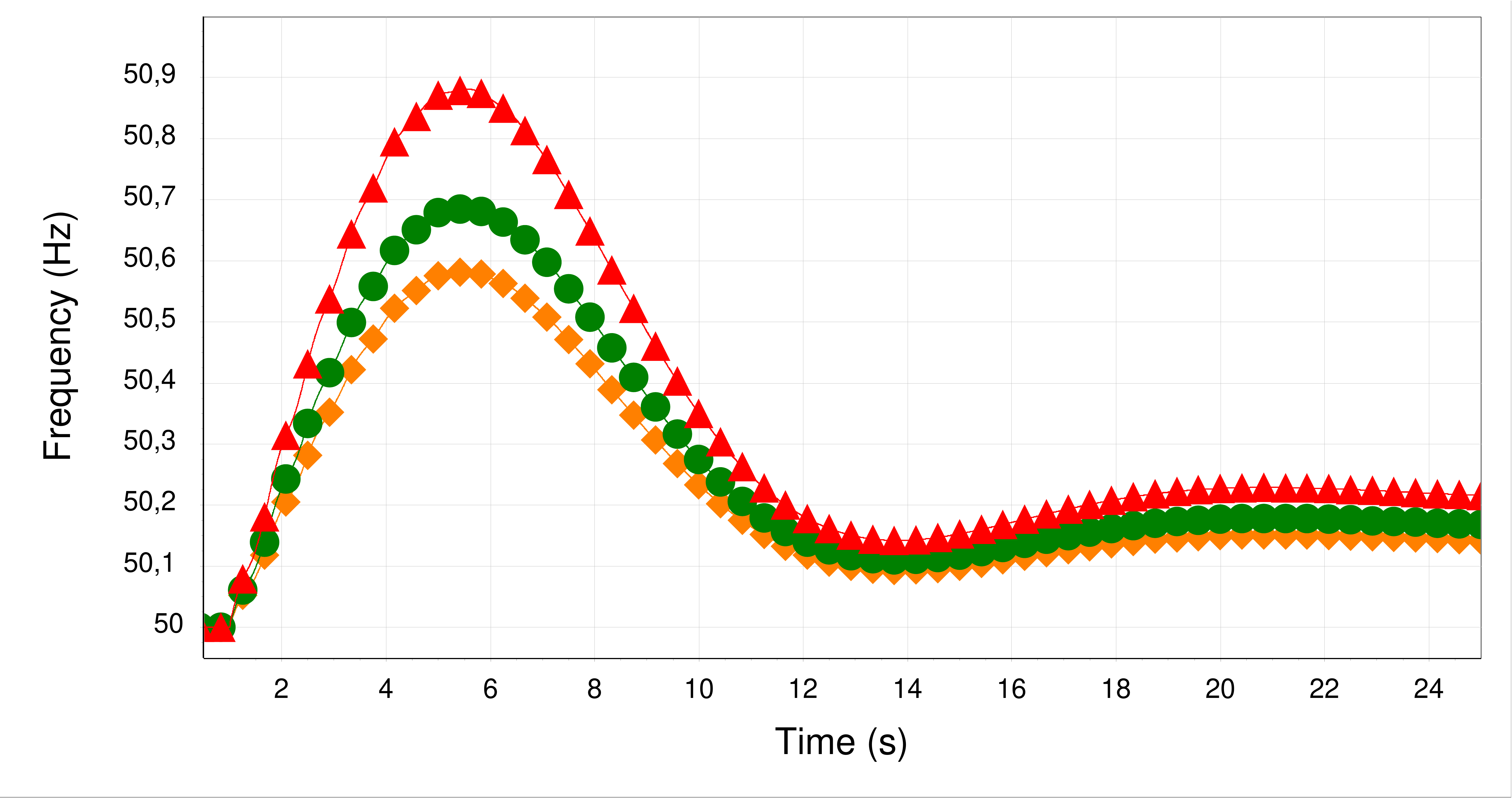}
        \put(65,50){\tikz\draw[red,fill=red] (0,0) -- (0.22,0) -- (0.11,0.22) -- cycle; 12\% reduced demand}
        \put(65,45){\tikz\draw[ao,fill=ao] (0,0) circle (.7ex); 9.4\% reduced demand}
        \put(65,40){\tikz\draw[rotate around={45:(0,0)},orange,fill=orange] (0,0) rectangle (0.2,0.2); 8\% reduced demand}
    \end{overpic}
    \caption{Frequency response of three DR attack scenarios with varying demand reductions.}
    \label{figure:pia_static_incidents}   
\end{figure}
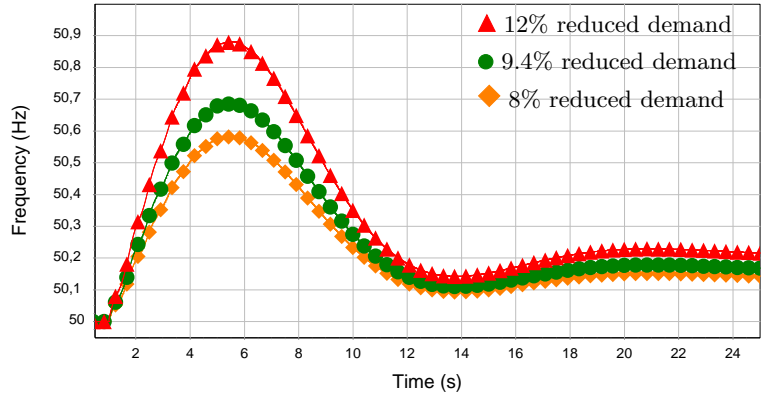

Since DR attacks are the opposite of DI attacks, the frequency response of such attacks should be inverted. To compare the differences between these, the three DI attacks with capacities corresponding to real incidents, as presented in Figure \ref{figure:lca_static_bus8}, are converted into DR attacks. The magnitudes of the DR attacks are thus 8\%, 9.4\%, and 12\% of the total generation. The impact of these attacks on the grid's frequency is presented in Figure \ref{figure:pia_static_incidents}. 

The incidents described in Section~\ref{cha:background} involve a sudden decrease in power supply to the grid, making them valuable benchmarks for evaluating DI and SR attacks. However, there have been no incidents involving large power injections or sudden reductions in demand within the Swedish power grid. Thus, it is not possible to compare DR attacks with any actual incidents. Instead, comparisons between DI and DR attacks are performed. 

The frequency response presented in Figure \ref{figure:pia_static_incidents} demonstrates that DR attacks have an inverted frequency response compared to DI attacks and that the magnitude of the frequency response for DR attacks is slightly larger than that of DI attacks. Reducing demand by 12\% of the total generation in the grid results in a peak frequency of 50.88 Hz (a deviation of 0.88 Hz), while increasing demand by 12\% caused the frequency to drop to 49.17 Hz (0.83 Hz deviation). This suggests that the grid is slightly more vulnerable and sensitive to DR attacks. This behavior aligns with Sweden's grid characteristics, where down-regulation reserves have less capacity and slower response times than reserves providing up-regulation, as presented in Table \ref{table:fcr}.

\begin{figure}[htbp]
    \centering
    \begin{subfigure}[b]{0.49\textwidth}
        \centering
        \begin{overpic}[width=\linewidth,clip,trim=0cm 0.3cm 0.3cm 0cm]{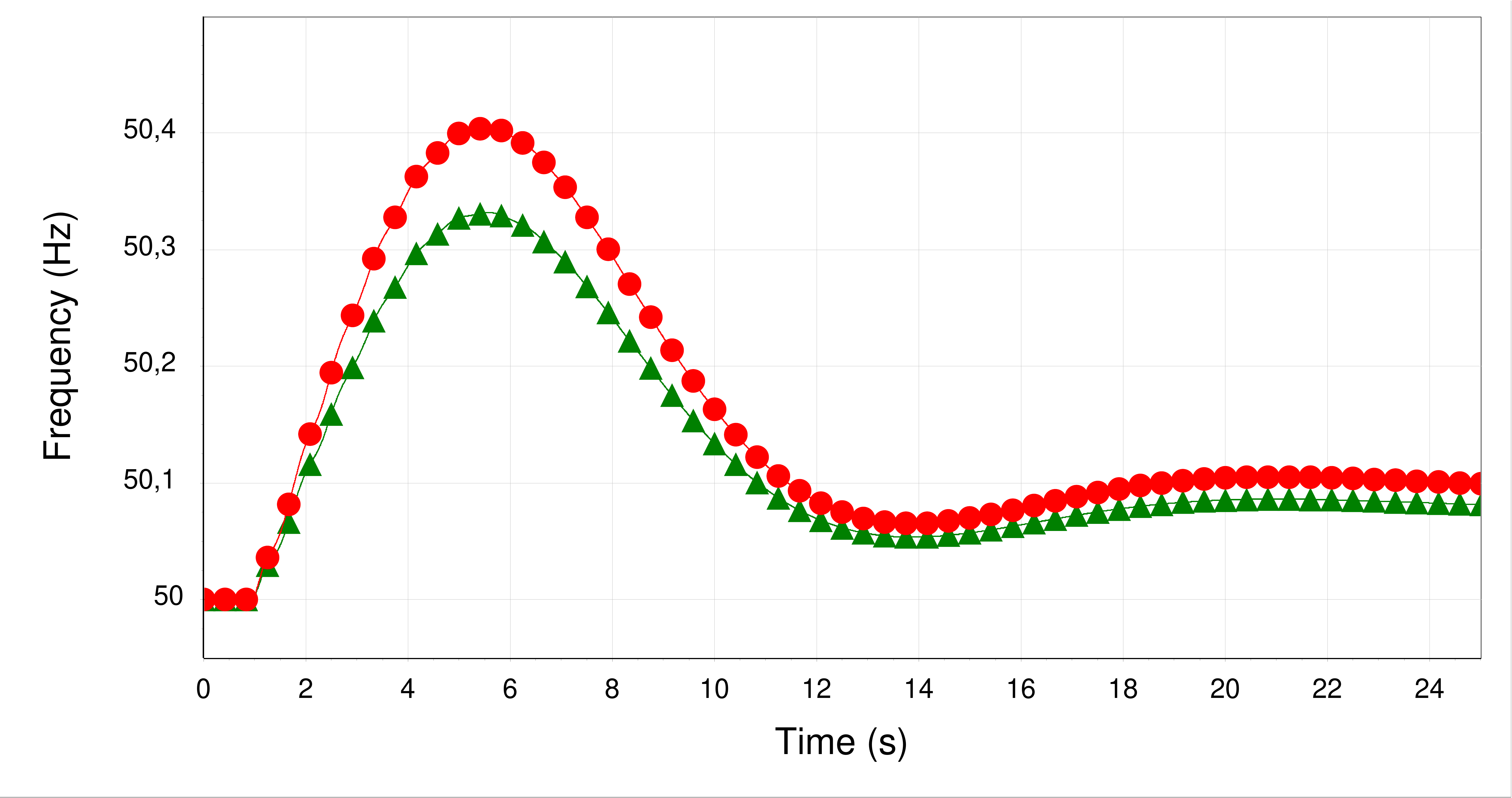}
            \put(80,98){\tikz\draw[red,fill=red] (0,0) circle (.7ex); 5.6\% reduced demand (2025)}
            \put(80,88){\tikz\draw[ao,fill=ao] (0,0) -- (0.22,0) -- (0.11,0.22) -- cycle; 4.6\% reduced demand (2030)}
        \end{overpic}
        \caption{Frequency response without FCR mitigation.}
        \label{figure:pia_before_mechanisms}
    \end{subfigure}
    \hfill
    \begin{subfigure}[b]{0.49\textwidth}
        \centering
        \begin{overpic}[width=\linewidth,clip,trim=0cm 0.3cm 0.3cm 0cm]{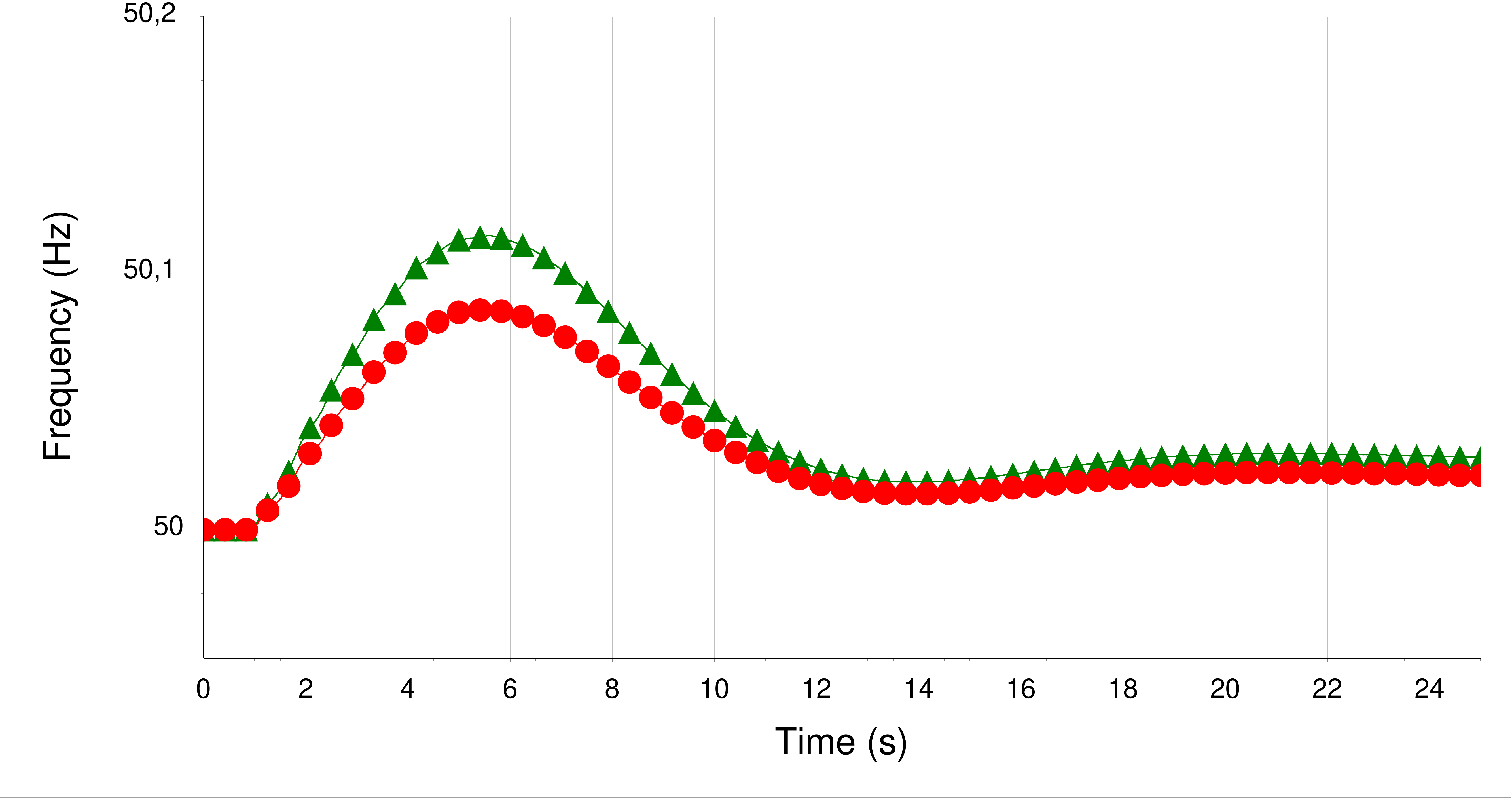}
            \put(78,98){\tikz\draw[red,fill=red] (0,0) circle (.7ex); 1.2\% reduced demand (2025)}
            \put(78,88){\tikz\draw[ao,fill=ao] (0,0) -- (0.22,0) -- (0.11,0.22) -- cycle; 1.6\% reduced demand (2030)}
        \end{overpic}
        \caption{Frequency response with FCR mitigation.}
        \label{figure:pia_after_mechanisms}
    \end{subfigure}
    \caption{Frequency response of two DR attacks.}
\end{figure}
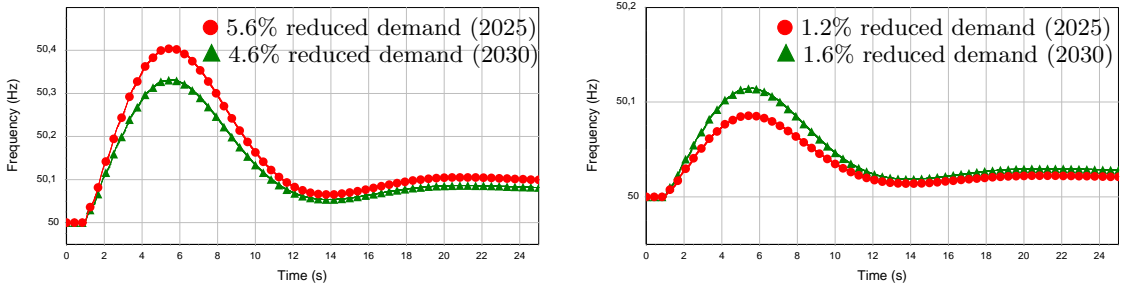

It is unlikely that attackers manage to control all the FERs. Instead, controlling a subset of them is more probable. For instance, FERs such as large-scale battery systems, are estimated to provide 1000 MW of flexibility in 2025 and 1200 MW in 2030, corresponding to 5.6\% and 4.6\%, respectively. By 2030, large-scale battery systems will represent a smaller share of total generation due to predicted growth in overall generation. Both 1000 MW and 1200 MW exceed the downregulation frequency reserves of FCR-Normal and downward FCR. Therefore, these frequency reserves are not sufficient to completely mitigate a DR attack of this size. DR attacks with these magnitudes are executed and presented in Figure~\ref{figure:pia_before_mechanisms}, which shows the frequency increasing to 50.40 Hz (2025) and 50.33 Hz (2030). Although these frequency deviations do not violate important frequency stability parameters, if these large-scale battery systems are concentrated within a small geographic area, their effect on the local grid could be more severe.

The combined capacity of FCR-Normal and downward FCR was 782 MW in 2024, which is not expected to increase significantly by 2026, as per the projections \cite{svk_future_reserves}. This paper assumes that the same capacities will be true for 2030, representing a worst-case scenario where the capacities of these frequency reserves are not increased. Assuming that these frequency reserves instantly and linearly respond to changes in demand and supply, the remaining magnitude of the 1000 MW DR attack is 218 MW and 418 MW for the 1200 MW DR attack. These remaining capacities correspond to 1.2\% in 2025 and 1.6\% in 2030 of the total generation. DR attacks of these magnitudes are executed and presented in Figure~\ref{figure:pia_after_mechanisms}. The result shows that the frequency remains within the normal operating range, indicating that DR of this size should not pose a significant challenge to grid stability, provided all FFR-Normal and downward FCR products are activated. 

\subsection*{Switching Attacks}
\label{switchingattacks}

Switching attacks are performed by executing a static DR, DI, SI, or SR attack followed by a sudden reversion to the original demand or supply at a predetermined time. Although relatively straightforward, this attack requires the attacker to have the capability to rapidly and synchronously change the power demand or supply of FERs. However, not all FERs can realistically change their demand or supply of energy quickly enough due to various limitations, such as ramp rates, that restrict the speed at which they can adjust their consumption or supply. In addition to the sudden imbalance caused by the switching attack, the activation of frequency reserves can be exploited to further amplify the imbalance between supply and demand. This is possible because frequency reserves respond to frequency deviations with a predictable delay. An attacker may alter the demand or supply of FERs so that they, during this time period, align with the frequency reserve's response.

\begin{figure}[htbp]
    \centering
    \begin{subfigure}[b]{0.49\textwidth}
        \centering
        \begin{overpic}[width=\linewidth,clip,trim=0cm 0.3cm 0.3cm 0cm]{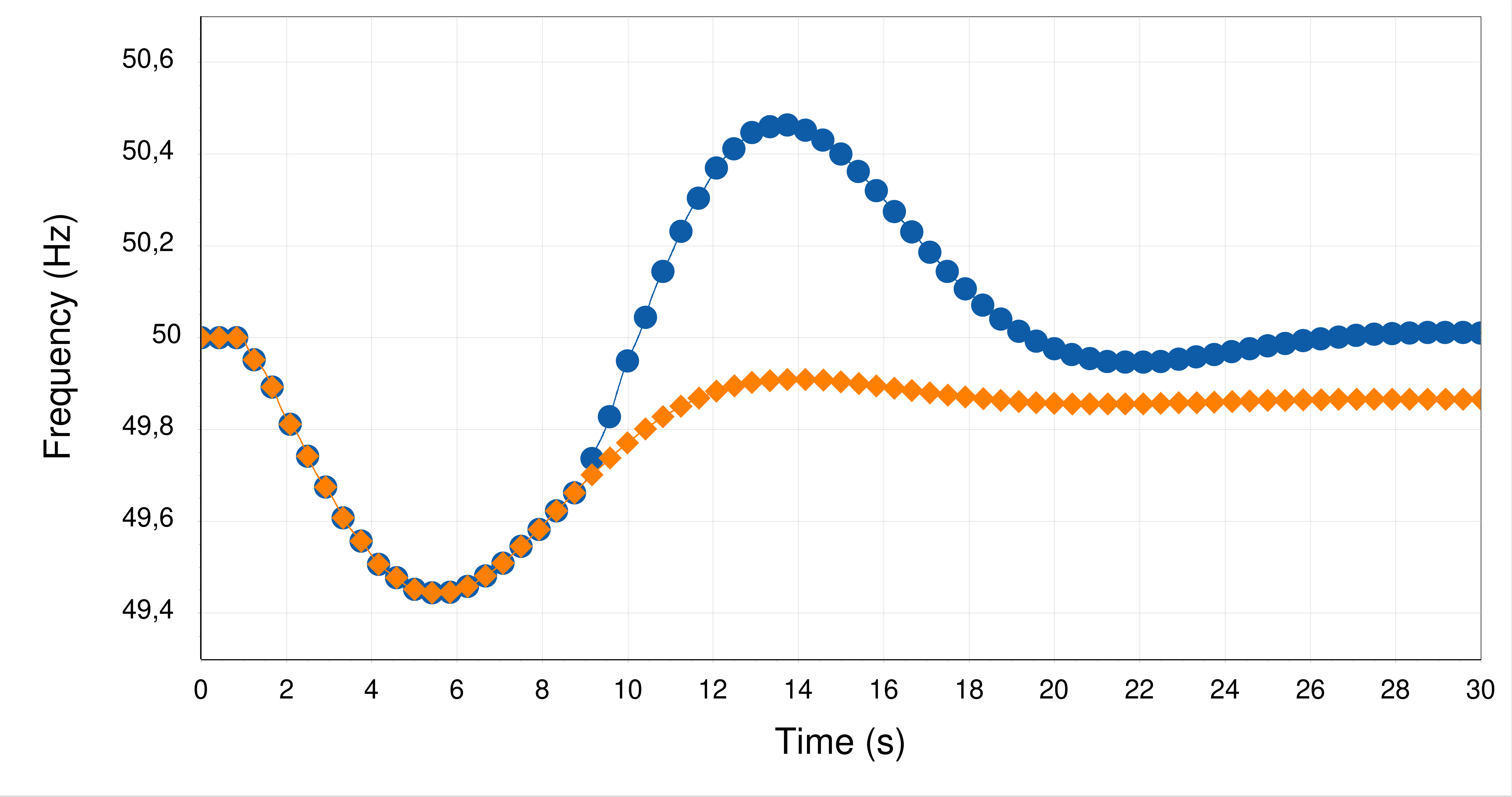}
            \put(90,35){\tikz\draw[aoblue,fill=aoblue] (0,0) circle (.7ex); DI switching attack (8\%)}
            \put(90,25){\tikz\draw[rotate around={45:(0,0)},orange,fill=orange] (0,0) rectangle (0.2,0.2); Static DI attack (8\%)}
        \end{overpic}
        \caption{Frequency response to a DI switching attack.}
        \label{figure:load_switching_comparison}
    \end{subfigure}
    \hfill
    \begin{subfigure}[b]{0.49\textwidth}
        \centering
        \begin{overpic}[width=\linewidth,clip,trim=0cm 0.3cm 0.3cm 0cm]{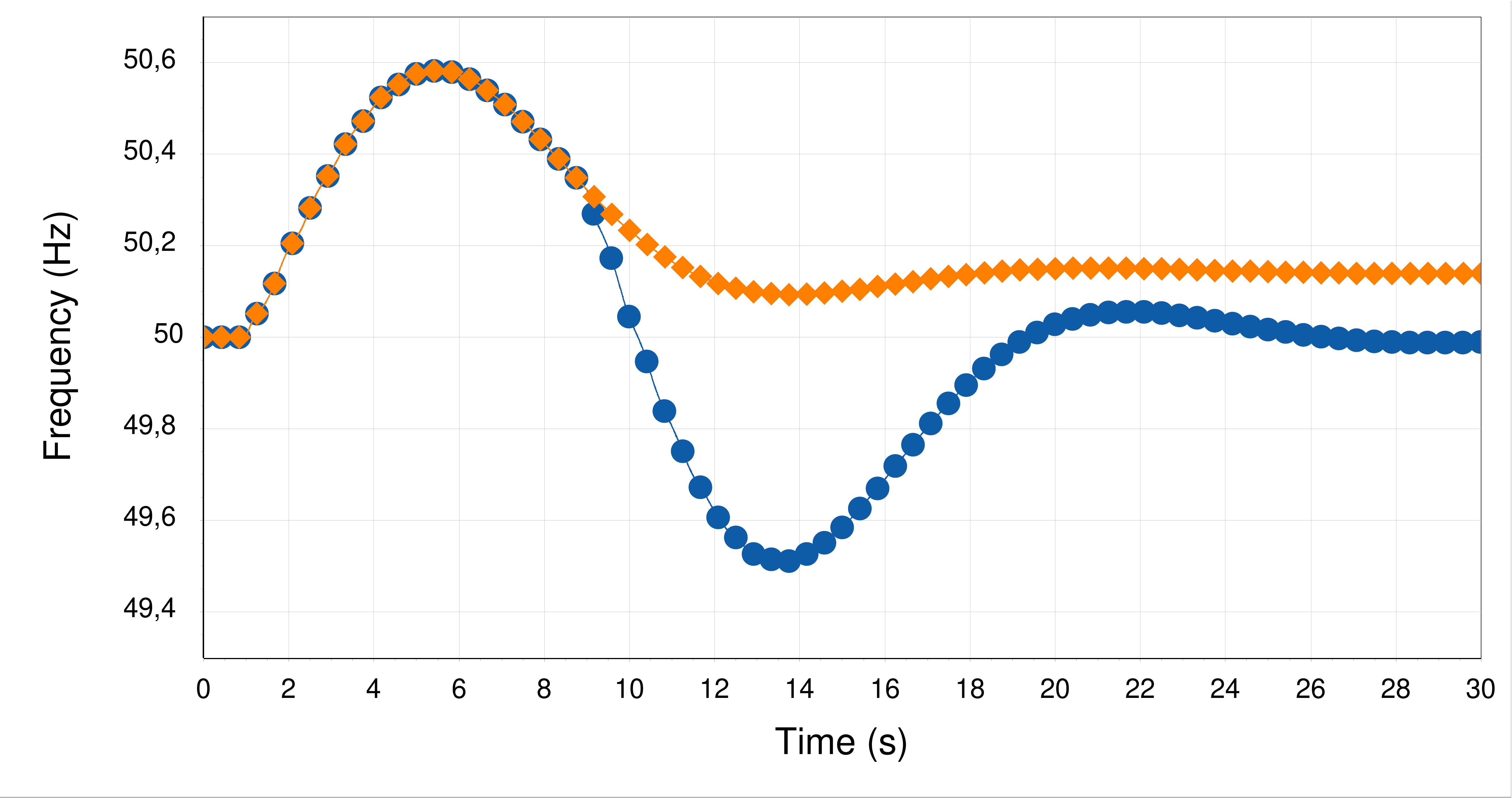}
            \put(90,95){\tikz\draw[aoblue,fill=aoblue] (0,0) circle (.7ex); DR switching attack (8\%)}
            \put(90,85){\tikz\draw[rotate around={45:(0,0)},orange,fill=orange] (0,0) rectangle (0.2,0.2); Static DR attack (8\%)}
        \end{overpic}
        \caption{Frequency response to a DR switching attack.}
        \label{figure:injection_switching_comparison}
    \end{subfigure}
    \caption{Impact of switching attacks on grid frequency.}
    \label{fig:combined_switching}
\end{figure}
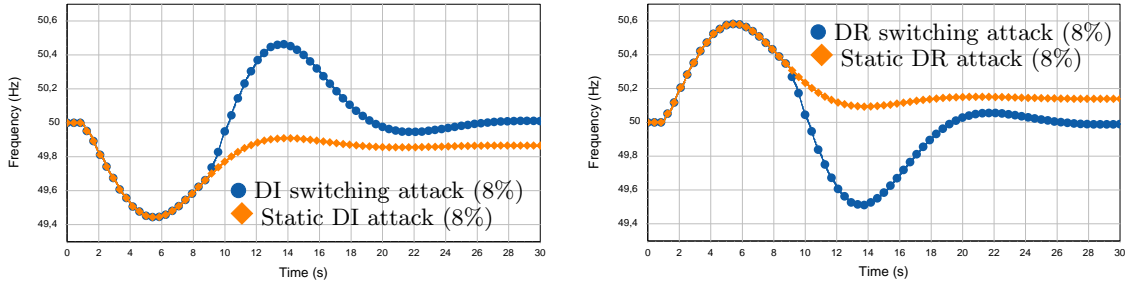

We examine two types of switching attacks. The first involves initially decreasing the frequency by performing a static DI or SR attack and then reverting it to the original demand or supply. The second type involves initially increasing the frequency through a DR or an SI attack, followed by the reversion. A DI-switching attack is executed and presented in Figure \ref{figure:load_switching_comparison}, while a DR attack is shown in Figure \ref{figure:injection_switching_comparison}, where each switching attack increases or decreases demand by 8\%. For comparison, static DI and DR attacks are also displayed in the respective figures. Each switching attack changes the demand by 8\% of the initial total supply in the grid. The static attack is executed after one second and then continues for seven seconds, at which point the reversion occurs. The precise timing of the reversal is determined to be critical for maximizing the disturbance to grid stability.

The figures above illustrate the impact of switching attacks. The attacks are essentially static attacks with an additional reversion component, and are thus identical up to the point of the reversion to the original demand. Thereafter, the sudden reversion back to the original demand causes the frequency to rapidly increase in Figure \ref{figure:load_switching_comparison} while rapidly decreasing in Figure \ref{figure:injection_switching_comparison}. Since the demand is returned to the original, the frequency ultimately stabilizes at the nominal frequency range. Comparing these results with their static counterparts, it is evident that these do not result in a higher or lower frequency than their equivalent static attacks. However, the quick change in frequency, from 49.4 to 50.4 Hz, causes oscillation and is very disruptive to the stability of the grid. It could also potentially impact voltage and phase stability of the grid. 

\begin{figure}[htbp]
    \centering
    \begin{subfigure}[b]{0.49\textwidth}
        \centering
        \begin{overpic}[width=\linewidth,clip,trim=0cm 0.3cm 0.3cm 0cm]{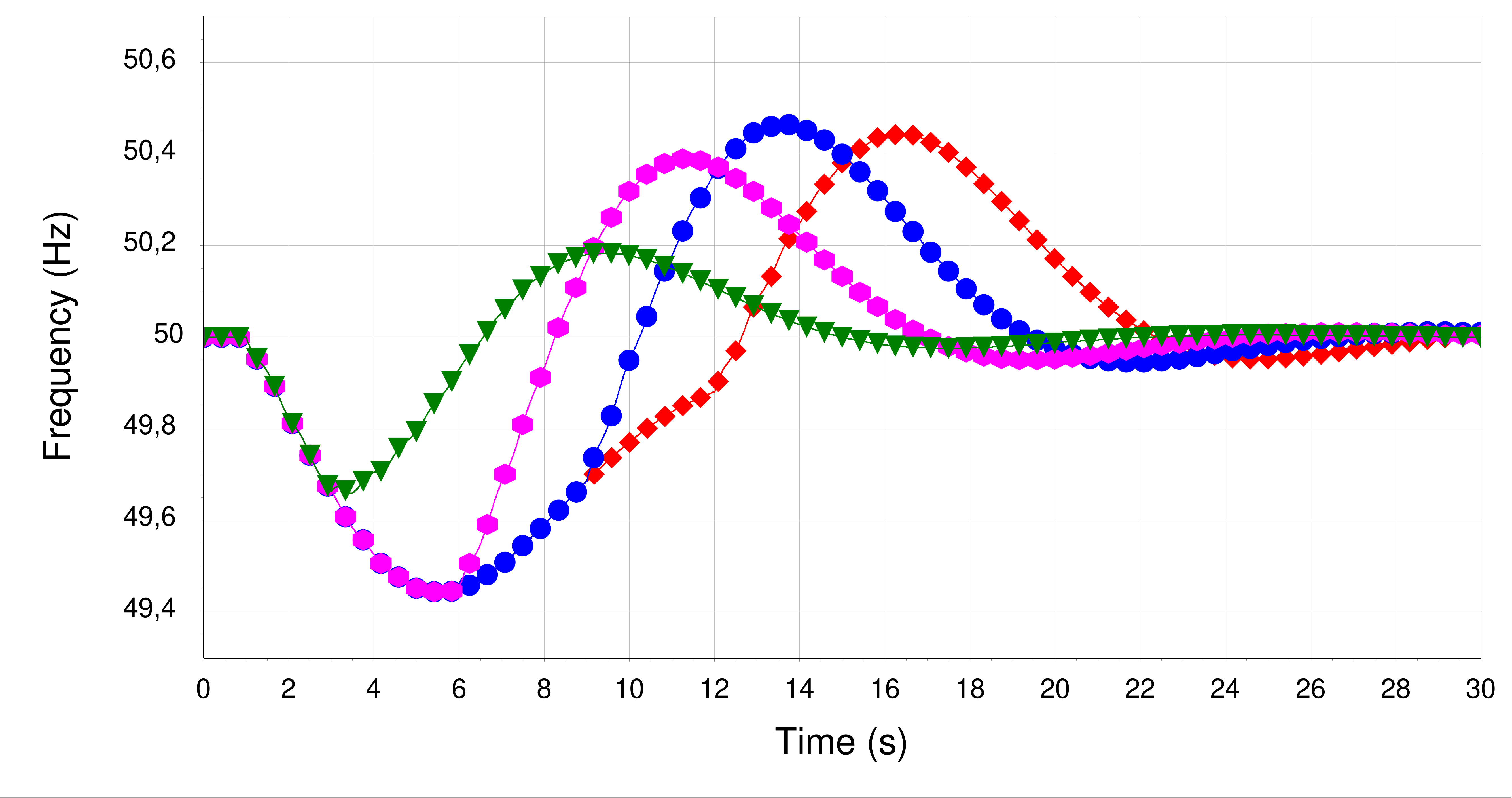}
            \put(87,45){\tikz\draw[ao,fill=ao] (0,0) -- (0.22,0) -- (0.11,0.22) -- cycle; \scriptsize DI switching attack at $T_1=3$}
            \put(86,37){\tikz\node[regular polygon,draw,regular polygon sides = 6, fill=pinkpink, inner sep=2.2pt] (p) at (0,0) {}; \scriptsize DI switching attack at $T_1=6$}
            \put(86,30){\tikz\draw[blueblue,fill=blueblue] (0,0) circle (.7ex); \scriptsize DI switching attack at $T_1=9$}
            \put(86,22){\tikz\draw[rotate around={45:(0,0)},red,fill=red] (0,0) rectangle (0.2,0.2); \scriptsize DI switching attack at $T_1=12$}
        \end{overpic}
        \caption{Four DI switching attacks with reversal times at $T_1=3$, $T_1=6$, $T_1=9$, and $T_1=12$ (s).}
        \label{figure:lca_diff_t}
    \end{subfigure}
    \hfill
    \begin{subfigure}[b]{0.49\textwidth}
        \centering
        \begin{overpic}[width=\linewidth,clip,trim=0cm 0.3cm 0.3cm 0cm]{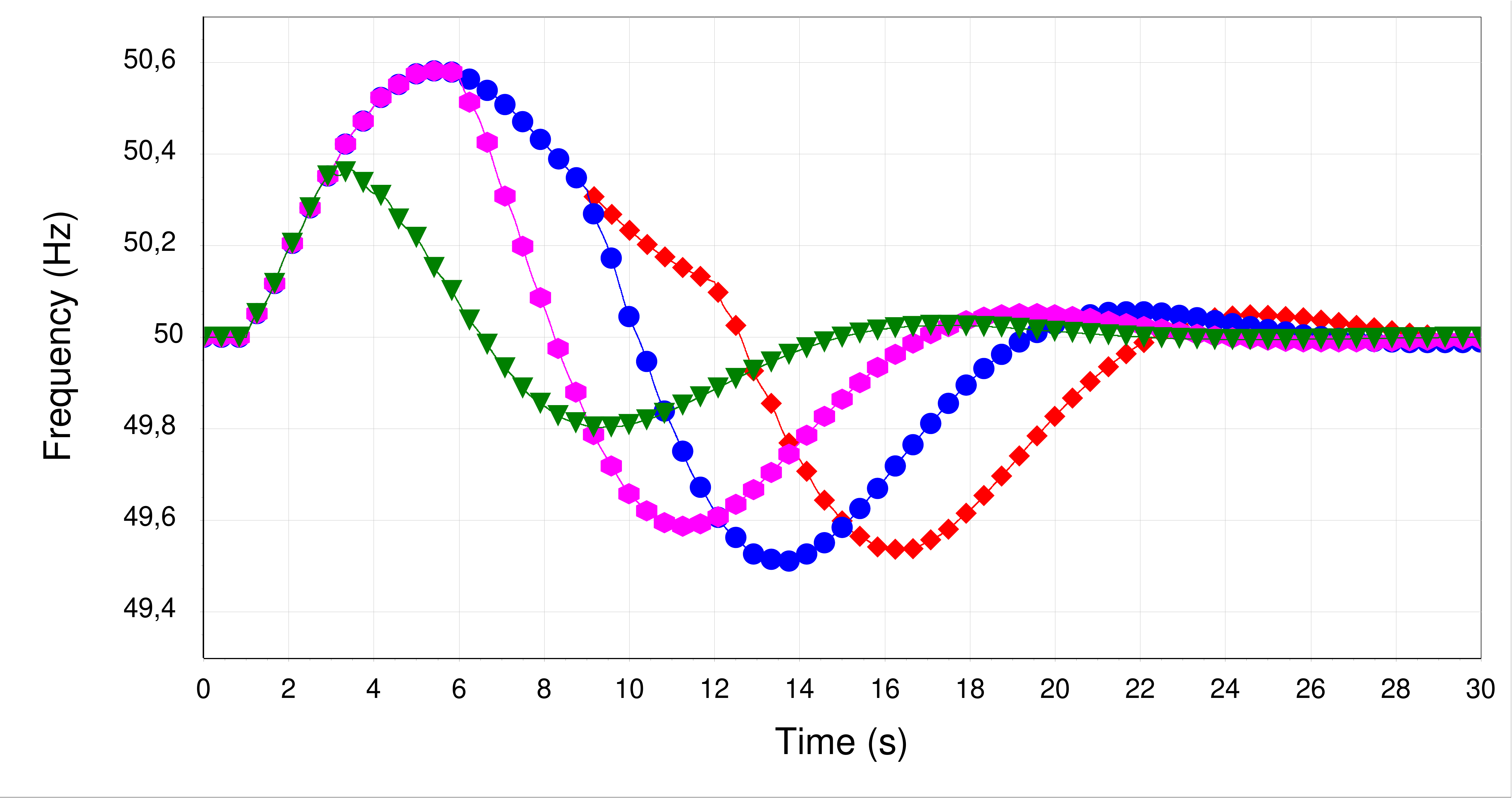}
            \put(87,102){\tikz\draw[ao,fill=ao] (0,0) -- (0.22,0) -- (0.11,0.22) -- cycle; \scriptsize DR switching attack at $T_1=3$}
            \put(87,95){\tikz\node[regular polygon,draw,regular polygon sides = 6, fill=pinkpink, inner sep=2.2pt] (p) at (0,0) {}; \scriptsize DR switching attack at $T_1=6$}
            \put(87,88){\tikz\draw[blueblue,fill=blueblue] (0,0) circle (.7ex); \scriptsize DR switching attack at $T_1=9$}
            \put(85,80){\tikz\draw[rotate around={45:(0,0)},red,fill=red] (0,0) rectangle (0.2,0.2); \scriptsize DR switching attack at $T_1=12$}
        \end{overpic}
        \caption{Four DR switching attacks with reversal times at $T_1=3$, $T_1=6$, $T_1=9$, and $T_1=12$ (s).}
        \label{figure:pia_diff_t}
    \end{subfigure}
    \caption{Impact of different reversal times ($T_1$) in switching attacks.}
\end{figure}
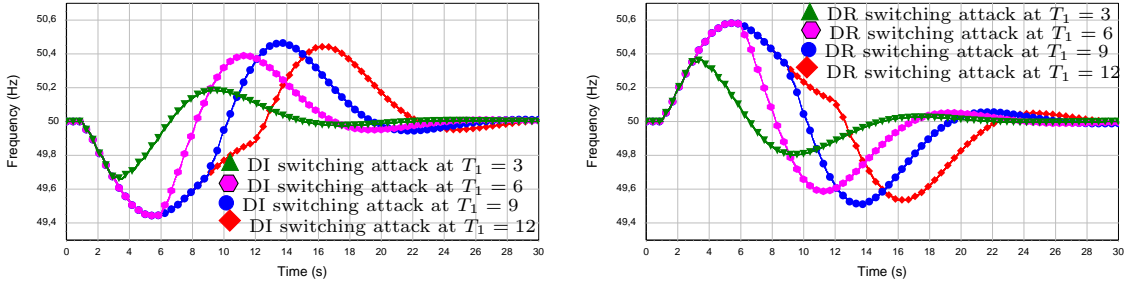

The timing of the reversion in switching attacks, where the FERs revert back to the original demand, is critical for maximizing the disturbance to grid stability. This is because the grid's frequency reserves have a delayed response. To investigate this, several switching attacks with various reversion timings are executed. Four DI-switching attacks with reversion occurring at different times (3, 6, 9, and 12 seconds) are shown in Figure \ref{figure:lca_diff_t}. Similarly, Figure \ref{figure:pia_diff_t} illustrates DR-switching attacks with the same reversion times.

The presented DI and DR attacks illustrate how varying the reversal time ($T_1$) affects the grid's frequency response. In this grid model, the optimal time for the reversion component of the switching attack is about 8 seconds. These results indicate that switching attacks do not achieve larger deviations from the nominal frequency than achieved by executing static DI, DR, SI, and DR attacks. However, switching attacks that induce sudden changes in demand and supply could still disrupt the grid's stability as they can repeatedly swing the grid frequency and cause oscillations. 

\subsection*{Periodic Attacks}
\label{sec:eval_periodicattacks}
Periodic attacks are executed by performing a series of switching attacks at specific intervals, with some duration between each attack. In this approach, once one switching attack is executed, the next switching attack is performed after a certain duration. This sequence creates a continuous chain of switching attacks that together form a periodic attack. As each switching attack involves a reversion, the grid is subjected to repeated fluctuations, leading to continuous changes in the demand or generation of FERs. 

Previous observations of switching attacks demonstrate that the duration between the switching attacks is important for maximizing their impact on grid stability. Thus, this timing has a profound effect on periodic attacks. To investigate this further, three periodic attacks are executed in total, each consisting of a series of four DI-switching (8\% increase) attacks with varying durations between each attack. The frequency responses from periodic attacks with 4, 8, and 16 seconds between each switching attack are presented in the figures below. 

\begin{figure}[htbp]
    \centering
    \begin{subfigure}[b]{0.49\textwidth}
        \centering
         \includegraphics[width=\linewidth, trim={0cm 0cm 15cm 0cm}, clip]{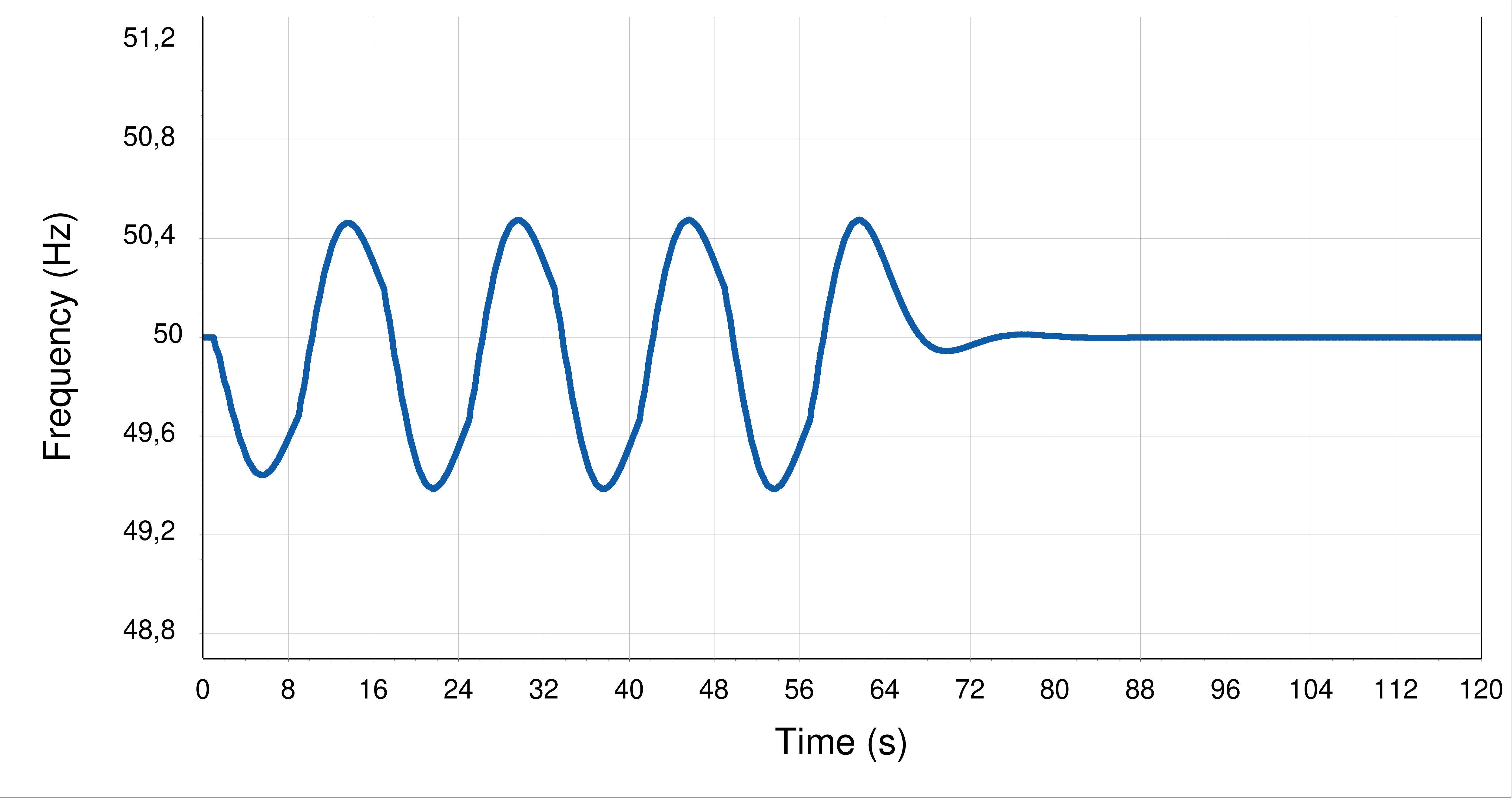}
         \caption{Periodic attack with 8 seconds intervals.}
        \label{figure:alternating2}
    \end{subfigure}
    \hfill
    \begin{subfigure}[b]{0.49\textwidth}
        \centering
        \includegraphics[width=\linewidth, trim={0cm 0cm 15cm 0cm}, clip]{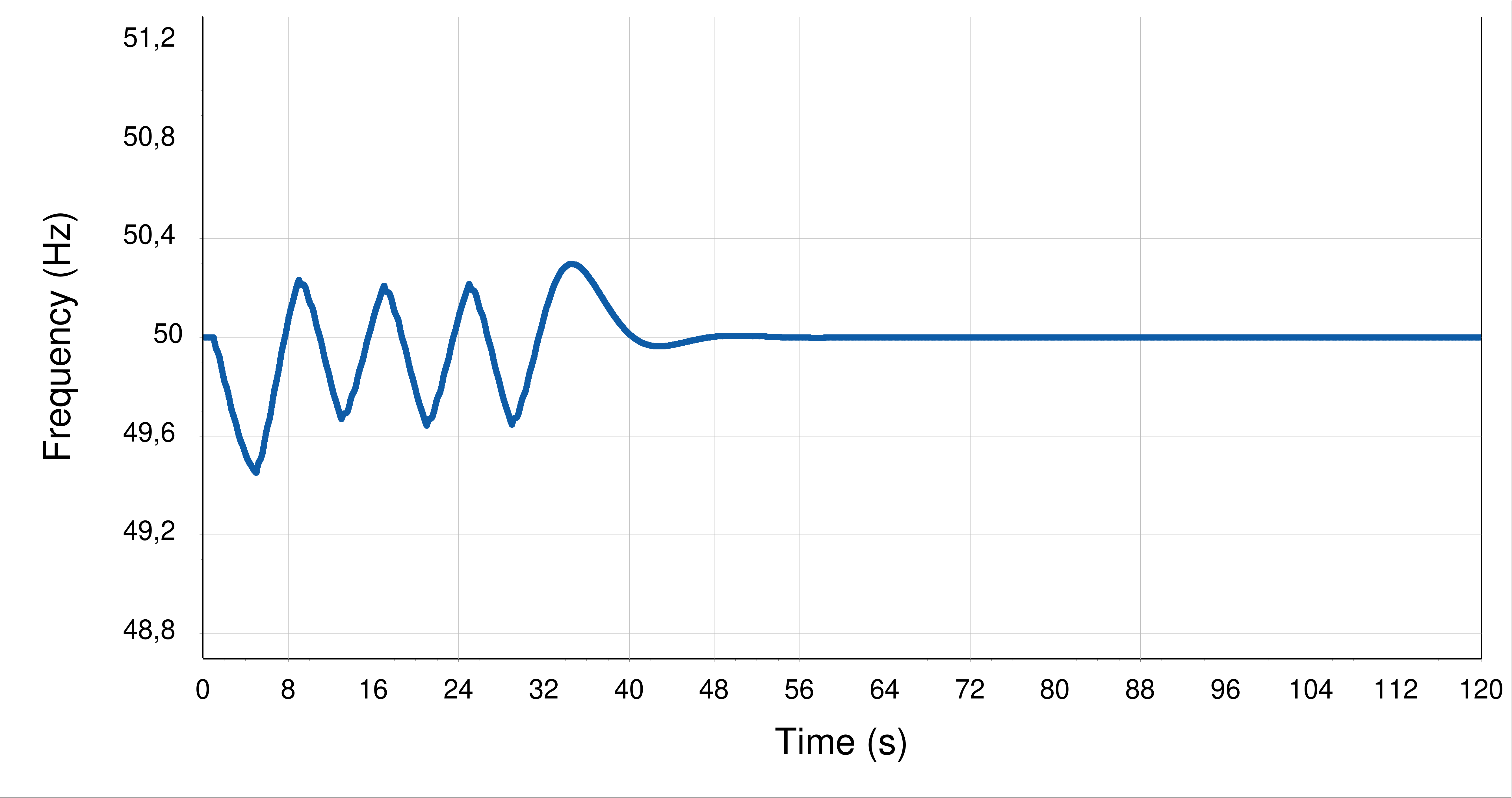}
        \caption{Periodic attack with 4 seconds intervals.}
        \label{figure:alternating1}
    \end{subfigure}
    \caption{Impact of periodic attacks on grid frequency.}
\end{figure}

\begin{figure}[htbp]
    \centering
    \begin{subfigure}[b]{0.49\textwidth}
        \centering
        \includegraphics[width=\linewidth]{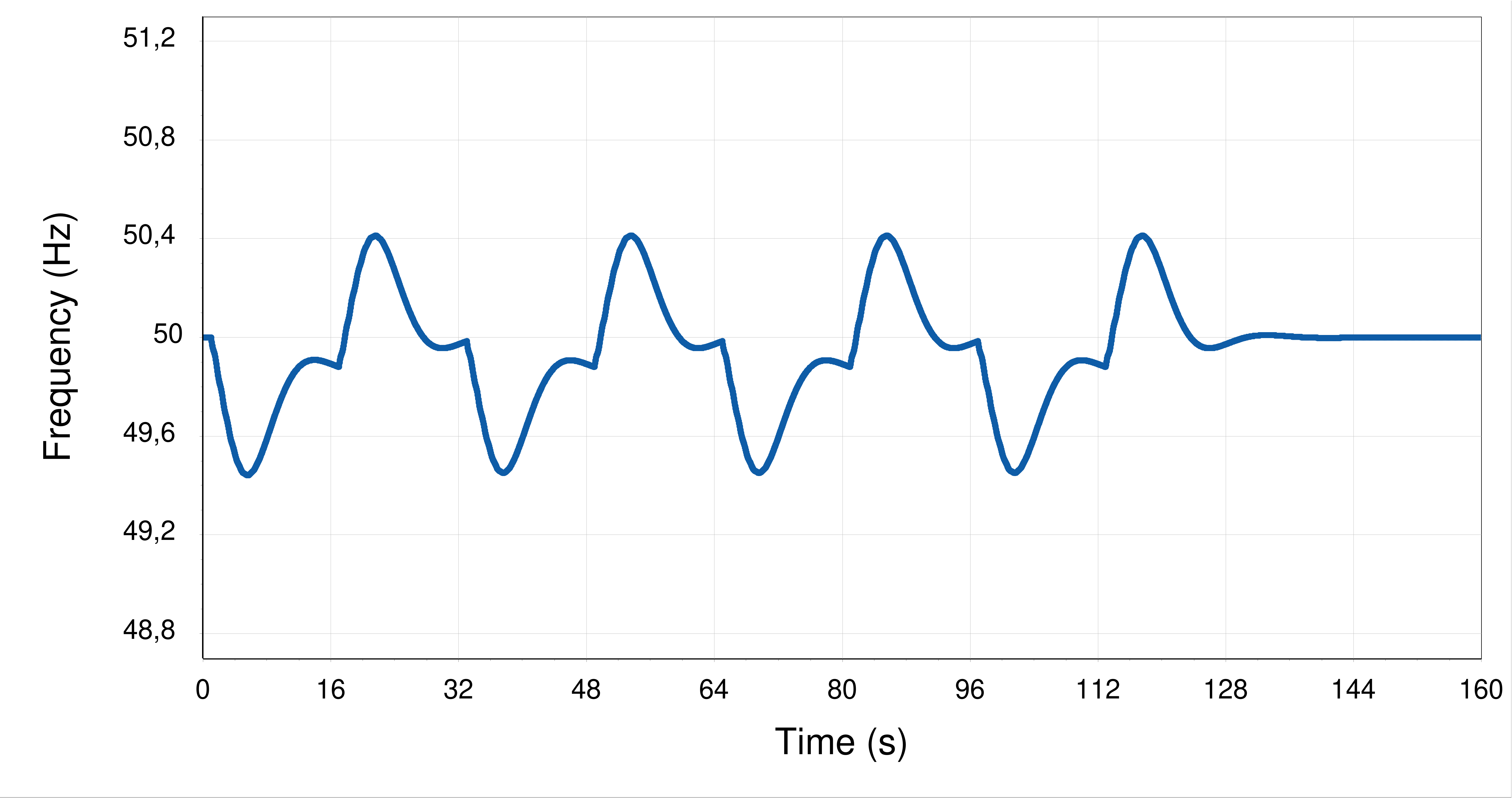}
        \caption{Periodic attack with 16 seconds intervals.}
        \label{figure:alternating3}
    \end{subfigure}
    \hfill
    \begin{subfigure}[b]{0.49\textwidth}
        \centering
        \includegraphics[width=\linewidth]{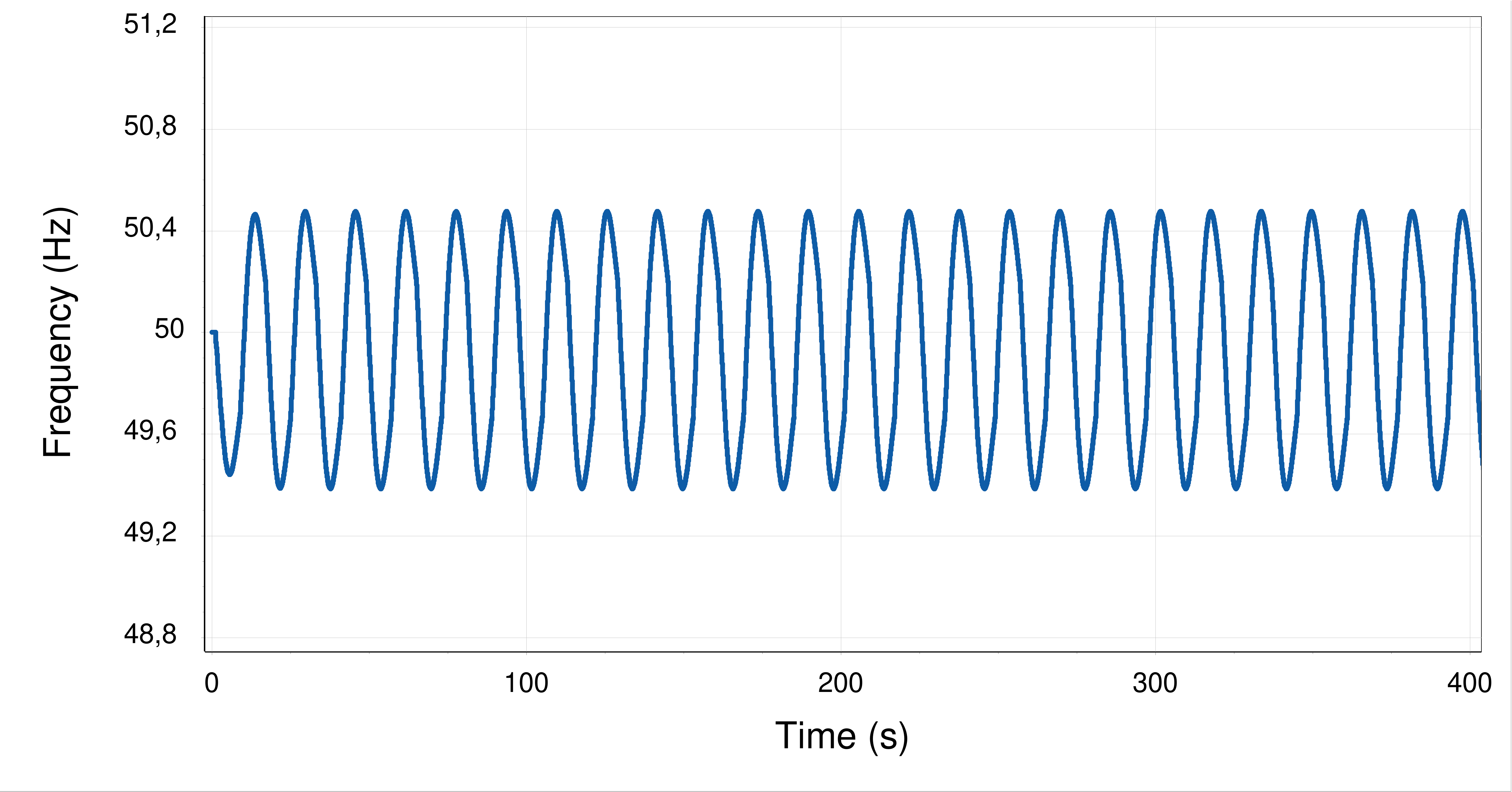}
        \caption{Long periodic attack with 50 DI-switching events.}
       \label{figure:alternating_400}
    \end{subfigure}
    \caption{Impact of periodic attacks of varying duration and quantity.}
\end{figure}

A periodic attack with a duration of 8 seconds between the switching attacks is presented in Figure \ref{figure:alternating2}, which reflects the optimal timing for switching attacks as described earlier. The intervals of 4 and 16 seconds are chosen to explore both half and double the optimal timing, providing a broader understanding of how different durations between the switching attacks, in a periodic attack, affect grid stability. The frequency responses for periodic attacks with 4 and 16-second intervals are illustrated in Figures \ref{figure:alternating1} and \ref{figure:alternating3}, respectively.

The periodic attacks cause the frequency to oscillate around the nominal frequency. Comparing the frequency response, in Figure \ref{figure:alternating2}, of the periodic attack with 8 seconds between each switching attack with Figures \ref{figure:alternating1} and \ref{figure:alternating3}, it is evident that 8 seconds between each switching attack induces the largest frequency deviation and is thus more effective from an attacker's perspective. In addition, periodic attacks do not induce a larger frequency deviation than regular switching attacks. 

Further examination of the frequency response in Figure \ref{figure:alternating2} indicates that the frequency deviation slightly increases over the entire attack duration. The highest achieved frequency during this periodic attack is 50.476 Hz at approximately 62 seconds, compared to 50.465 Hz in Figure \ref{figure:load_switching_comparison} at roughly 14 seconds. Building upon this observation, a longer periodic attack lasting 400 seconds and consisting of 50 switching attacks is executed and presented in Figure~\ref{figure:alternating_400}. For this periodic attack, an 8-second interval between each switching attack is chosen, as it has been identified as the most optimal duration for achieving the largest frequency deviation in this model. However, the results in Figure \ref{figure:alternating_400} indicate that the frequency does not consistently grow over time, with the highest measured frequency being 50.476 Hz at 93 seconds. In addition, a similar attack performed over a duration of $10,000$ seconds produced similar results, indicating that frequency deviations do not grow over time. This suggests that the periodic nature of the attacks does not lead to increasingly large deviations in frequency.

Due to their continuous nature, periodic attacks might be more challenging to mitigate than static and switching attacks. Mitigating a periodic attack would involve continuously activating and deactivating the frequency reserves to counteract the imbalance. Without effective mitigation, periodic attacks sustained over time could significantly challenge the grid's ability to operate within normal operating ranges. The resulting frequency oscillations could place stress on critical grid infrastructure  such as connected loads and generators, potentially leading to damage or causing them to disconnect from the grid. If such disconnection were to occur, there is an increased risk of cascading failures throughout the power grid as more generators or loads may need to be disconnected. Another challenge is identifying the assets involved, as many FERs are potentially distributed across a large geographical area, and each one only constitutes a small portion of the total attack.

\subsection*{Combination Attacks}

The definition of combination attack allows for mixed directions of switching attacks, with the first attack increasing the frequency and the subsequent attack decreasing it, or vice versa. By evaluating this combined attack, this paper aims to investigate whether this attack has a worse impact on the grid's stability and if a more complex attack strategy is more successful in exploiting the frequency response pattern from a sequence of attacks. 

\begin{figure}[htbp]
    \centering
    \begin{subfigure}[b]{0.49\textwidth}
    \centering
    \includegraphics[width=\linewidth]{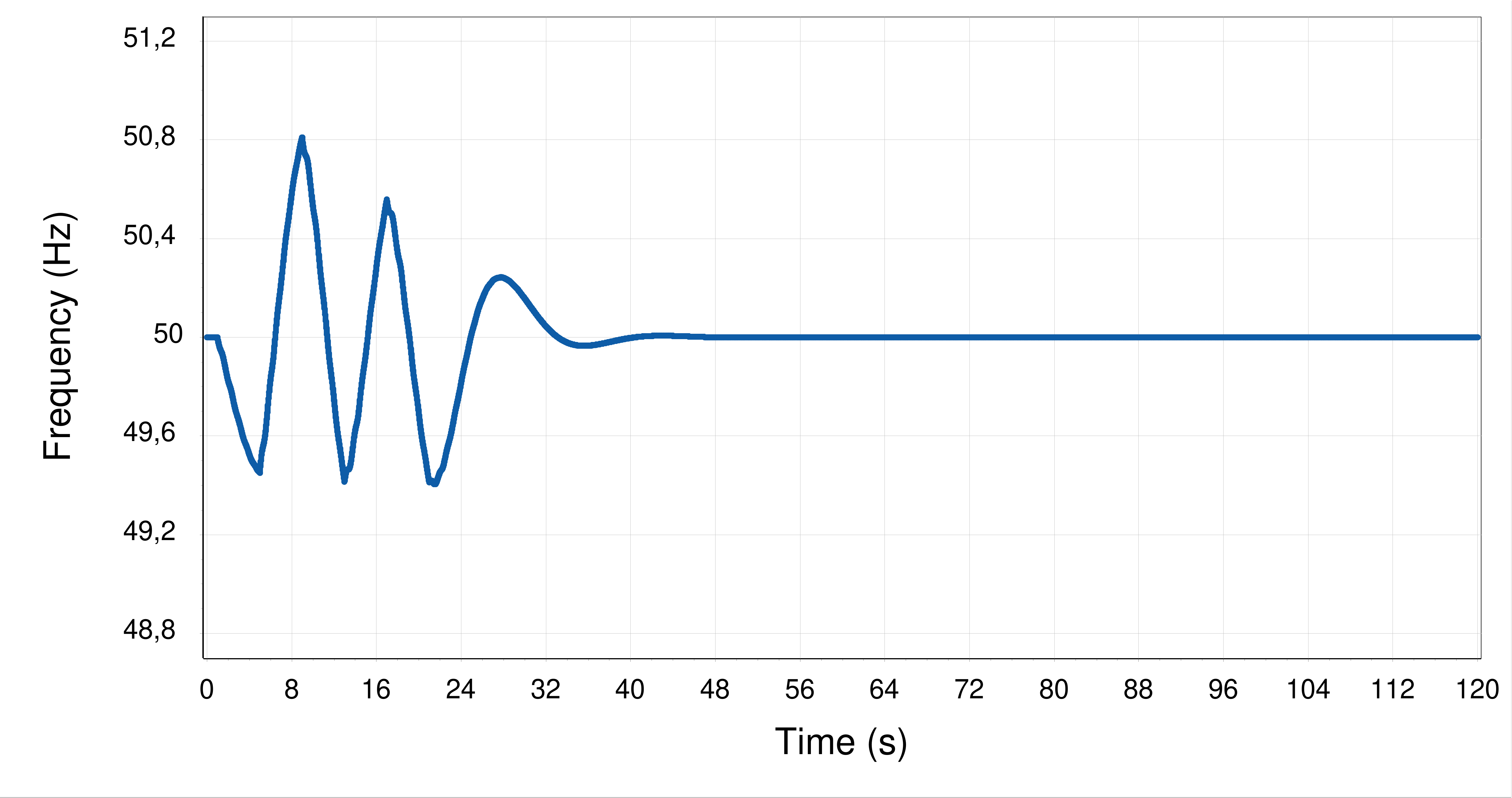}
    \caption{Combination attack with 4 seconds interval.}
    \label{figure:dynamic_sequence_1}
    \end{subfigure}
    \hfill
    \begin{subfigure}[b]{0.49\textwidth}
        \centering
         \includegraphics[width=\linewidth]{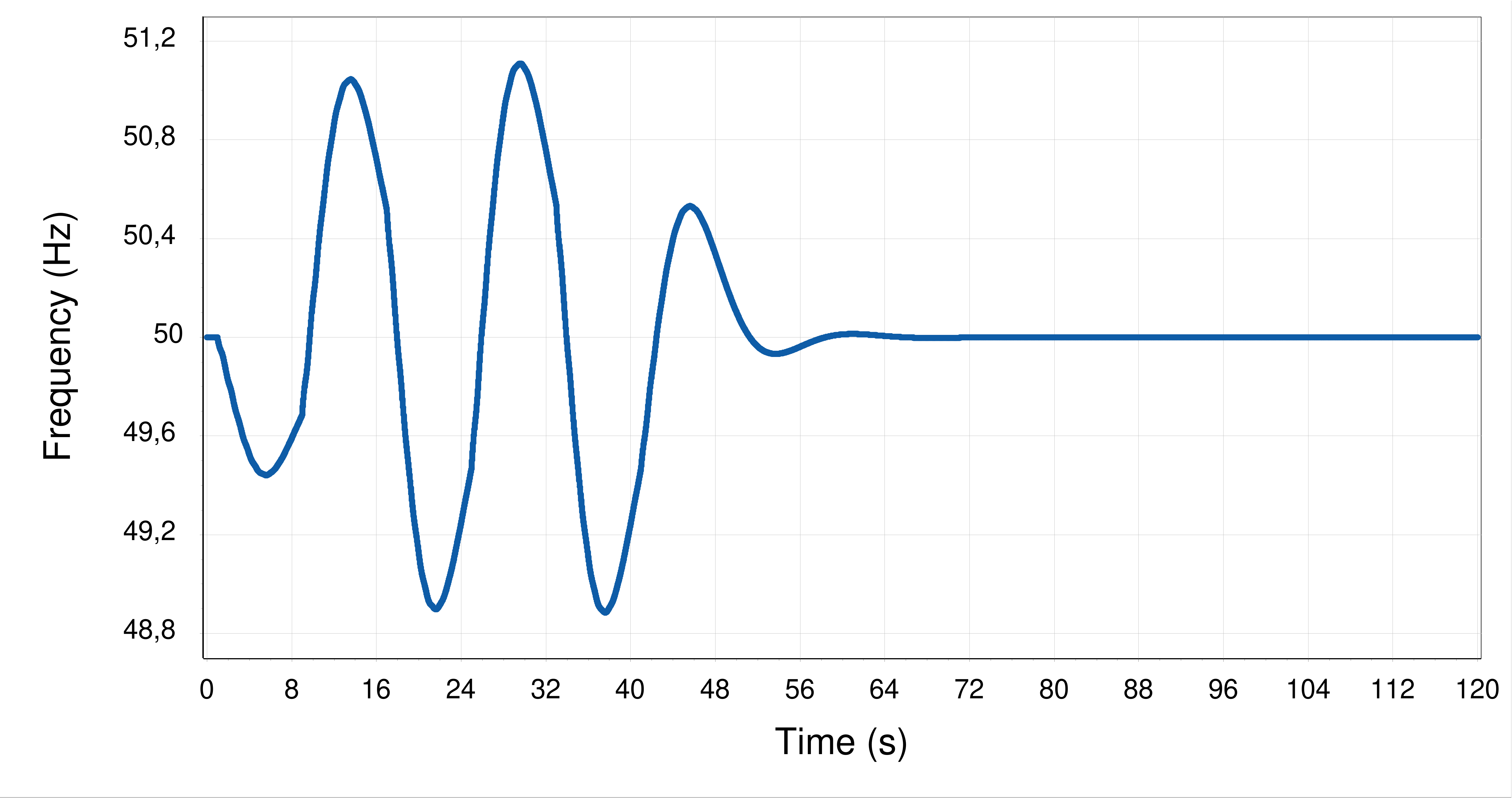}
        \caption{Combination attack with 8 seconds interval.}
        \label{figure:dynamic_sequence_2}
    \end{subfigure}
    \caption{Impact of combination attacks on grid frequency.}
\end{figure}

\begin{figure}[htbp]
    \centering
    \begin{subfigure}[b]{0.49\textwidth}
        \centering
        \includegraphics[width=\linewidth]{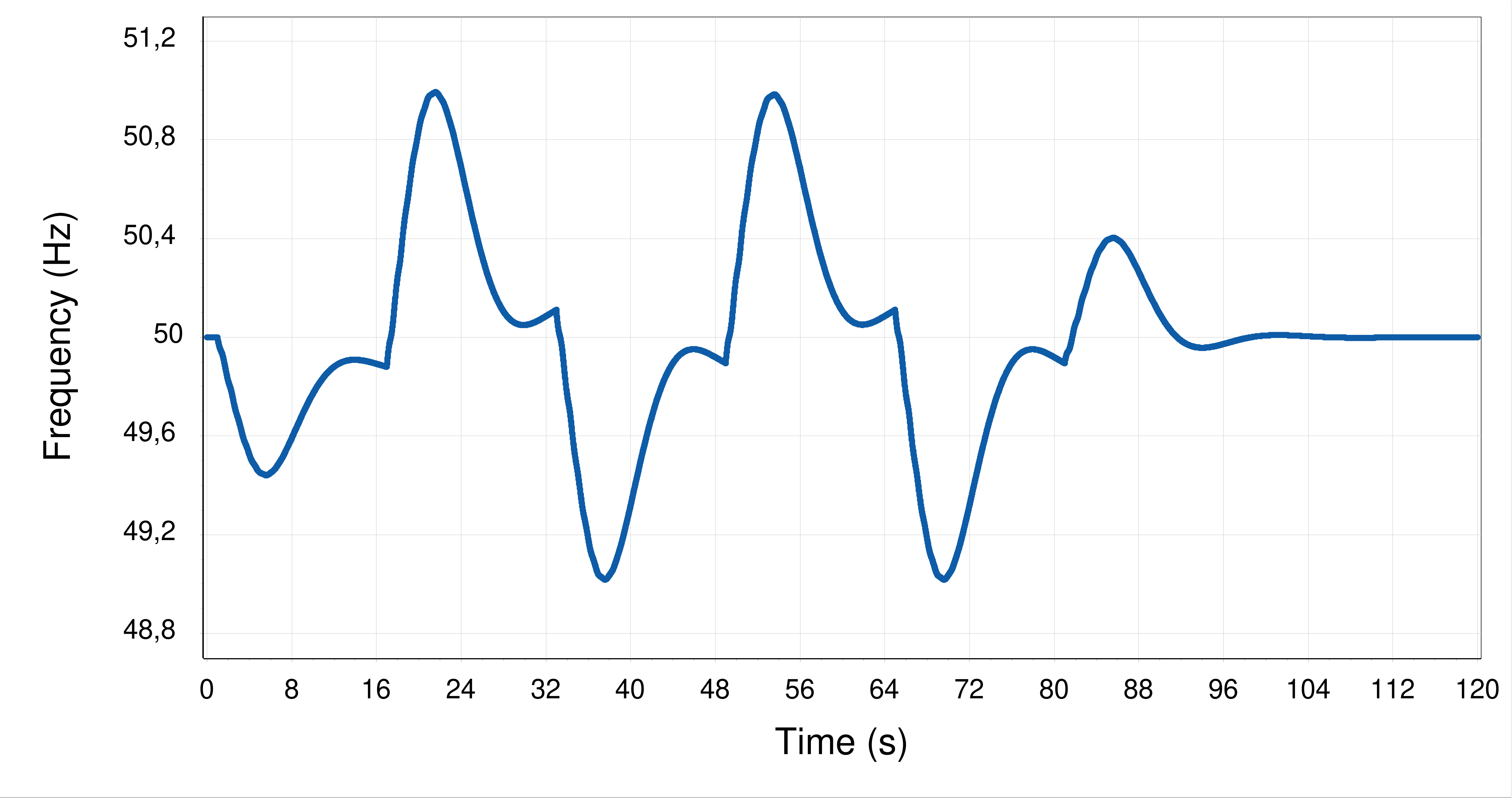}
        \caption{Combination attack with 16 seconds interval.}
        \label{figure:dynamic_sequence_3}
    \end{subfigure}
    \hfill
    \begin{subfigure}[b]{0.49\textwidth}
        \centering
        \begin{overpic}[width=\linewidth,clip, trim=0cm 0.3cm 0.3cm 0cm]{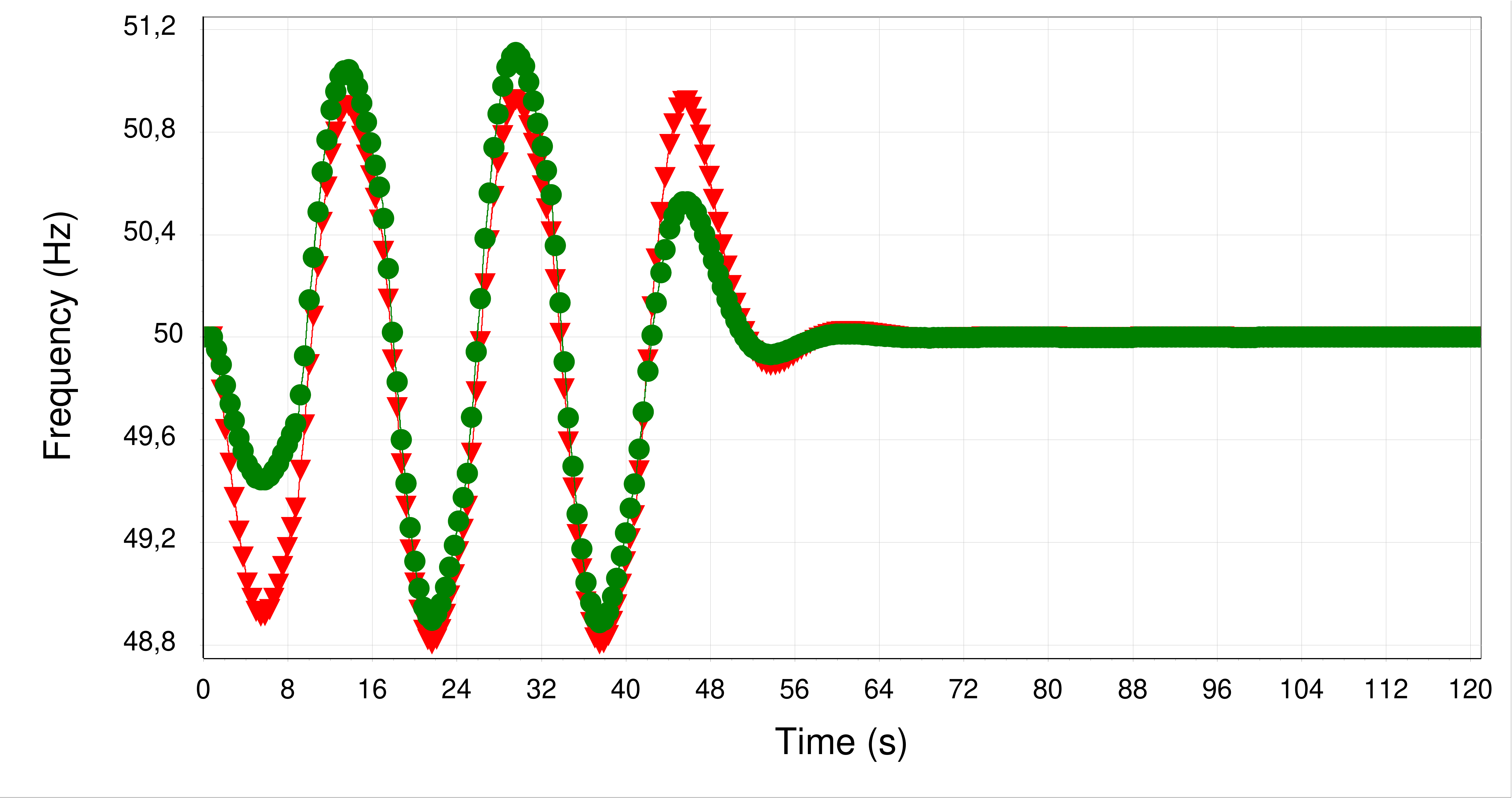}
            \put(95,47){\tikz\draw[red,fill=red] (0,0) -- (0.22,0) -- (0.11,0.22) -- cycle; periodic attack (16\%)}
            \put(95,37){\tikz\draw[ao,fill=ao] (0,0) circle (.7ex); combination attack (8\%)}
        \end{overpic}
        \caption{Comparison of combination and periodic attack.}
        \label{figure:dynamic_sequence_4}
    \end{subfigure}
    \caption{Impact of combinations attacks of varying intervals and equivalent periodic attack.}
\end{figure}

Three combination attacks that alternate between increasing demand by 8\% and subsequently decreasing it by 8\% are presented in the following figures. Each combination attack employs a different timing interval, specifically 4, 8, or 16 seconds, based on previous evaluations and to ease comparison. The results for these combination attacks are presented in Figure \ref{figure:dynamic_sequence_1} (4 seconds), Figure \ref{figure:dynamic_sequence_2} (8 seconds), and Figure \ref{figure:dynamic_sequence_3} (16 seconds).

The impact of the combined attacks as shown in figures is similar to previous periodic attacks, but the corresponding frequency deviation is more than twice as large from the nominal frequency compared to periodic attacks. The highest achieved frequency for combination attacks is 51.11 Hz, which can be observed in Figure \ref{figure:dynamic_sequence_2}, while the best periodic attack, from an attacker's perspective, achieved the highest frequency of 50.476 Hz. A larger impact for combination attacks is not surprising, as the alternating pattern effectively doubles the magnitude of the attack, first increasing demand by 8\%, then returning it to the original demand, and finally decreasing it by 8\%. Consequently, combination attacks that incorporate both DI and DR attacks are capable of causing larger frequency deviations and thus larger disturbances to first-swing stability than other attacks, that either increase or reduce demand, but not both simultaneously.

Since the combination attack effectively doubles the attack magnitude, it is also of interest to see how it compares to periodic attacks that increase demand by similar amounts. A period attack increasing demand by 16\% (double that of previously evaluated periodic attacks) is presented in Figure  \ref{figure:dynamic_sequence_4}, alongside the frequency response of the combination attack demonstrated in Figure \ref{figure:dynamic_sequence_2} (8-second interval). The periodic attack that increases demand by 16\% achieves a similar frequency response as a combination attack that alternates between increasing and decreasing demand by 8\%. Therefore, it can be concluded that combination attacks can achieve the same impact as periodic attacks while only requiring changes in demand by half the amount. This is particularly interesting because the total change (16\%) for combination attacks remains the same as for periodic attacks. Moreover, the combination attack keeps the demand and supply levels closer to normal amounts which is likely to avoid tripping fuses, as well as voltage and current relays. 

From an attacker's perspective, a strategy that optimizes impact while evading detection and tripping of defenses is highly desirable. However, this strategy also comes with a caveat that the controlled FERs must be capable of executing both DI and DR switching attacks. This could be achieved by e.g., BESSs that first charge and then suddenly discharge, or through heat pumps that can initially increase their demand, restore it to original demand levels, and subsequently reduce demand by an equivalent amount. To conclude, combination attacks pose a greater risk to grid stability, while requiring fewer or smaller FERs to achieve the same impact as other attacks.

\section{Interpretation of Results}
\label{cha:discussion}

The aim of this paper is to identify and assesses potential cyberattack strategies targeting FERs in order to assess their impact on the first-swing stability of the power grid. A detailed evaluation of the attack scenarios is presented in Section~\ref{cha:evaluation} where the necessary steps and parameters for each scenario, including various time intervals and attack sizes, are provided, to ensure applicability and reliability. However, the evaluated attacks are only a representative selection of possible attack strategies and there might be other attacks that are not evaluated in this work. The experiments demonstrate that the impact of the different attack scenarios on the frequency in the power grid follows a linear relationship with the size of the induced change. For the static DI, DR, SI, and SR attacks, results demonstrate a distinct linear correlation between the lowest respective highest frequency and the size of the static attacks. According to the Swedish TSO, the regulation provided by the frequency reserves should be linear to the frequency deviation \cite{svk_linear_regulation}. This confirms the linear function derived in our results. Even though the attack scenarios' impact on the frequency follows a linear relationship, some are superior in other aspects. Periodic attacks, for example, can induce severe frequency oscillations that can potentially damage equipment and components connected to the grid while simultaneously disrupting grid stability. Whereas combination attacks that alternate between increasing and decreasing load can remain closer to normal operating values in terms of current and voltage while maximizing the frequency deviations, achieving a similar effect as periodic attacks but with half capacity. Other attacks, such as switching attacks, stand out due to being easy to execute, while still causing lots of disruption.

The timing of various attacks has also been shown to have a significant impact on the frequency stability of the power grid. To maximize frequency deviations, an attacker must identify the optimal timing of when to execute these attacks. For the simulated attacks, the optimal timing between attacks seems to be around eight seconds. However, the optimal timings identified in this paper are difficult to extrapolate to a real grid since the grid model recoups much faster than it would in reality. Nevertheless, the results suggest that the optimal timing for switching, periodic and combination attacks is when the frequency curve has the largest rate of change when approaching the nominal frequency. The power grid has shown some resilience to sudden and large static attacks, as demonstrated both in this work and in real-world incidents, where the grid remains stable and eventually returns to normal operating ranges. While the current capacities of FERs in 2025 pose a significant threat to the power grid on a national level they still require a substantial share of assets capable of providing flexibility. By 2030, the potential increase in FER capacities will allow attackers to more easily carry out attacks, both in terms of assets required and the potential attack surface. Moreover, the capacities of FERs are based on approximations of their flexibility potential, but an attacker could possibly take over other resources that could be remotely controlled but not currently used for flexibility, for example batteries in households. 

The presented work has an extensive focus on Sweden, utilizing its grid conditions, FER capacity, and frequency reserves. This scope allows for an in-depth analysis of potential attacks against a grid with realistic characteristics and provides an opportunity to assess whether existing frequency mitigation mechanisms can defend against the attacks. Nevertheless, our findings should be generally applicable to any representative grid with similar characteristics. To support replicability, configuration files for the simulator and attacks, are openly published on GitHub~\cite{danielmyren_powerworld_wscc9_configs}. While the presented work provides critical insights into aggregated attack strategies against the power grid using remotely connected FERs, several directions remain for future research. One idea could be to focus on enhancing the realism and complexity of the power grid model by incorporating geographically distributed generation and loads, line capacity constraints, frequency reserves, and protection mechanisms such as relays. These additions would allow for more accurate assessments of attack impacts, including cascading failures. Further research could also evaluate attack strategies under worst-case conditions, such as peak demand periods or adverse weather, where the grid is most vulnerable. Furthermore, the development of detection mechanisms remains critical. Future efforts could focus on identifying potential indicators of compromise within FER aggregators and applying frameworks like MITRE ATT\&CK to design robust detection and mitigation strategies.
\section{Conclusions}
\label{cha:conclusion}

This paper investigates the threats posed by adversaries who can gain control over multiple FERs within the power grid. The paper focuses on identifying potential aggregated attack strategies, assessing their impact on the grid's frequency stability, and evaluating the effectiveness of existing frequency reserves in mitigating these impacts. The work also incorporates realistic estimations of FER capacities to assess the risk of severe consequences, both today and in the future. The results show that all explored attack strategies impact the first-swing stability of the power gird. However, combination attacks are the most severe, capable of causing the largest frequency deviations, while controlling the same amount of power. The work demonstrates that such attacks are already feasible today in 2025 based on the current capabilities of FERs and represent a significant threat to the power grid. As FERs continue to grow in both prevalence and capacity in the coming years, the associated risks and threats are expected to escalate, highlighting the urgent need for robust security measures and effective frequency mitigation strategies.


\begin{credits}
\subsubsection{\ackname} The work was conducted as part of the Cybersecurity for Resilient Energy Communities of the Future (CyREC) project, funded by the Swedish Innovation Agency (ref nr. 2023-02987). 
\end{credits}

\bibliographystyle{splncs04}
\bibliography{references}

\begin{thebibliography}{10}
\providecommand{\url}[1]{\texttt{#1}}
\providecommand{\urlprefix}{URL }
\providecommand{\doi}[1]{https://doi.org/#1}

\bibitem{acharya_load_alt_ev}
Acharya, S., Dvorkin, Y., Karri, R.: Public plug-in electric vehicles + grid data: Is a new cyberattack vector viable? IEEE Transactions on Smart Grid  \textbf{11}(6),  5099--5113 (2020)

\bibitem{amini_load_alt_sim}
Amini, S., Pasqualetti, F., Mohsenian-Rad, H.: Dynamic load altering attacks against power system stability: Attack models and protection schemes. IEEE Transactions on Smart Grid  \textbf{9}(4) (2018)

\bibitem{inspection_bess}
Baumgart, I., Borsig, M., Goerke, N., Hackenjos, T., Rill, J., Wehmer, M.: Who controls your energy? on the (in)security of residential battery energy storage systems. In: IEEE International Conference on Communications, Control, and Computing Technologies for Smart Grids. pp.~1--6 (2019)

\bibitem{security_issue_ami}
Cleveland, F.M.: Cyber security issues for advanced metering infrasttructure (ami). In: 2008 IEEE Power and Energy Society General Meeting - Conversion and Delivery of Electrical Energy in the 21st Century. pp.~1--5 (2008)

\bibitem{iva_svangmassa}
D.~Karlsson, G.~Power, A.N.: Svängmassa i elsystemet. Tech. rep., Kungliga Ingenjörsvetenskapsakademien (2016)

\bibitem{dabrowski_load_altering}
Dabrowski, A., Ullrich, J., Weippl, E.R.: Grid shock: Coordinated load-changing attacks on power grids: The non-smart power grid is vulnerable to cyber attacks as well. In: Proceedings of the 33rd Annual Computer Security Applications Conference. p. 303–314. Association for Computing Machinery (2017)

\bibitem{hawaii37bus}
{Electric Grid Test Case Repository}: Hawaii synthetic grid (2024), \url{https://electricgrids.engr.tamu.edu/hawaii40/}, accessed: 2025-11-11

\bibitem{wscc9bus}
{Electric Grid Test Case Repository}: Wscc 9-bus system (2024), \url{https://electricgrids.engr.tamu.edu/electric-grid-test-cases/wscc-9-bus-system/}, accessed: 2025-11-11

\bibitem{nist_smart_grid}
Gopstein, A., Nguyen, C., O'Fallon, C., Hastings, N., Wollman, D.A.: Nist framework and roadmap for smart grid interoperability standards, release 4.0 (2021-02-18 00:02:00 2021)

\bibitem{scada_issues}
Igure, V.M., Laughter, S.A., Williams, R.D.: Security issues in scada networks. Computers \& Security  \textbf{25}(7),  498--506 (2006)

\bibitem{smartgrid_attacks_challenges}
Li, X., Liang, X., Lu, R., Shen, X., Lin, X., Zhu, H.: Securing smart grid: cyber attacks, countermeasures, and challenges. IEEE Communications Magazine  \textbf{50}(8),  38--45 (2012)

\bibitem{lin_false_data_injection}
Lin, J., Yu, W., Yang, X., Xu, G., Zhao, W.: On false data injection attacks against distributed energy routing in smart grid. In: Third International Conference on Cyber-Physical Systems (2012)

\bibitem{liu_false_data_injection}
Liu, Y., Ning, P., Reiter, M.K.: False data injection attacks against state estimation in electric power grids. ACM Transaction on Information and System Security  \textbf{14}(1) (2011)

\bibitem{simscape}
MathWorks: Simscape, \url{https://se.mathworks.com/products/simscape.html}

\bibitem{mohsenian_load_altering}
Mohsenian-Rad, A.H., Leon-Garcia, A.: Distributed internet-based load altering attacks against smart power grids. IEEE Transactions on Smart Grid  \textbf{2}(4),  667--674 (2011)

\bibitem{denai_v2g_impact}
Mojumder, M.R.H., Ahmed~Antara, F., Hasanuzzaman, M., Alamri, B., Alsharef, M.: Electric vehicle-to-grid (v2g) technologies: Impact on the power grid and battery. Sustainability  \textbf{14}(21) (2022)

\bibitem{problem_renewable_load_demand}
Moslehi, K., Kumar, R.: A reliability perspective of the smart grid. IEEE Transactions on Smart Grid  \textbf{1}(1),  57--64 (2010)

\bibitem{danielmyren_powerworld_wscc9_configs}
Myren, D.: The threat of aggregated cyberattacks on power grids (2025), \url{https://github.com/danielmyren/the-threat-of-aggregated-cyberattacks-on-power-grids/}

\bibitem{security-bess}
{\"{O}}hrstr{\"{o}}m, F., Oscarsson, J., Afzal, Z., Dani, J., Asplund, M.: From balance to breach: cyber threats to battery energy storage systems. Energy Inform.  \textbf{8}(1), ~39 (2025)

\bibitem{parks_ami}
Parks, R.C.: Advanced metering infrastructure security considerations. Tech. rep., Sandia National Laboratories (2007), \url{https://www.energy.gov/ceser/articles/advanced-metering-infrastructuresecurity-considerations}

\bibitem{powerworld}
PowerWorld: Powerworld simulator, \url{https://www.powerworld.com/products/simulator/overview}

\bibitem{1_der_smart_inverters}
Qi, J., Hahn, A., Lu, X., Wang, J., Liu, C.C.: Cybersecurity for distributed energy resources and smart inverters. IET Cyber-Physical Systems: Theory \& Applications  \textbf{1}(1),  28--39 (2016)

\bibitem{sayed_v2g_attacks}
Sayed, M.A., Atallah, R., Assi, C., Debbabi, M.: Electric vehicle attack impact on power grid operation. International Journal of Electrical Power \& Energy Systems  \textbf{137},  107784 (2022)

\bibitem{siemenspsse}
Siemens: Power system simulation, \url{https://www.siemens.com/global/en/products/energy/grid-software/planning/pss-software/pss-e.html}

\bibitem{soltan_load_alt_sim}
Soltan, S., Mittal, P., Poor, H.V.: {BlackIoT}: {IoT} botnet of high wattage devices can disrupt the power grid. In: 27th USENIX Security Symposium. pp. 15--32. USENIX Association (2018)

\bibitem{security_grid_state_of_art}
Sun, C.C., Hahn, A., Liu, C.C.: Cyber security of a power grid: State-of-the-art. International Journal of Electrical Power \& Energy Systems  \textbf{99},  45--56 (2018)

\bibitem{svk_linear_regulation}
{Svenska kraftnät}: Hur ska {FCR-reglering} ske i förhållande till frekvensen?, \url{https://www.svk.se/aktorsportalen/bidra-med-reserver/fragor-och-svar-om-reserver/fcr/hur-ska-fcr-reglering-ske-i-forhallande-till-frekvensen/}, accessed: 2025-11-11

\bibitem{svk_integrering_2013}
{Svenska kraftnät}: Integrering av vindkraft (2013), \url{https://www.svk.se/siteassets/om-oss/rapporter/2015-och-aldre/20130313-integrering-av-vindkraft.pdf}, accessed: 2025-11-11

\bibitem{svk_storning_2023}
{Svenska kraftnät}: Driftstörningen den 26 april 2023 (2023), \url{https://www.svk.se/siteassets/om-oss/rapporter/2023/rapport-driftstorning-2023-04-26_slutversion.pdf}, accessed: 2025-11-11

\bibitem{flex_report}
{Svenska kraftnät}: Främjandet av ett mer flexibielt elsystem (2023), \url{https://www.svk.se/siteassets/om-oss/rapporter/2023/framjande-av-ett-mer-flexibelt-elsystem-deluppdrag-5-ei-r2023-18.pdf}, accessed: 2025-11-11

\bibitem{svk_future_reserves}
{Svenska kraftnät}: Framtida volymbehov (2024), \url{https://www.svk.se/aktorsportalen/bidra-med-reserver/behov-av-reserver-nu-och-i-framtiden/framtida-volymbehov}, accessed: 2025-11-11

\bibitem{svk_reserves}
{Svenska Kraftnät}: Overview of requirements for reserves (2025), \url{https://www.svk.se/495bac/siteassets/aktorsportalen/bidra-med-reserver/om-olika-reserver/oversiktlig-kravbild-for-reserver-sve-20250324.pdf}, accessed: 2025-11-11

\bibitem{nordic_grid_plan}
{Svenska Kraftnät, Energinet, Fingrid, Statnett}: Nordic grid development perspective (2024), \url{https://www.svk.se/siteassets/om-oss/rapporter/2023/svk_ngpd2023.pdf}, accessed: 2025-11-11

\bibitem{un_ukraine}
{UN}: Attacks on {Ukraine’s} electricity infrastructure threaten key aspects of life, \url{https://ukraine.un.org/en/278995-attacks-ukraine’s-electricity-infrastructure-threaten-key-aspects-life-winter-approaches-–}, accessed: 2025-11-11

\bibitem{gevernova}
Vernova, G.: Steady state power flow, \url{https://www.gevernova.com/consulting/planos/steady-state-power-flow}, accessed: 2025-11-11

\bibitem{commmunication_grid}
Wang, W., Xu, Y., Khanna, M.: A survey on the communication architectures in smart grid. Computer Networks  \textbf{55}(15),  3604--3629 (2011)

\bibitem{xiang_aggregated}
Xiang, Y., Wang, L., Liu, N.: Coordinated attacks on electric power systems in a cyber-physical environment. Electric Power Systems Research  \textbf{149},  156--168 (2017)

\end{thebibliography}

\end{document}